\documentclass[acmlarge, 10pt]{acmart}

\usepackage{subcaption}
\usepackage{soul}
\usepackage{siunitx}
\AtBeginDocument{%
  }

\newcommand{\new}[1]{{\color{black}#1}}

\usepackage{balance}

\usepackage{xcolor}
\usepackage{ifthen}

\newboolean{editing}
\setboolean{editing}{true}  

\newcommand{\add}[1]{%
  \ifthenelse{\boolean{editing}}%
    {\textcolor{blue}{[Add: #1]}}%
    {#1}%
}

\newcommand{\delete}[1]{%
  \ifthenelse{\boolean{editing}}%
    {\textcolor{red}{[Delete: #1]}}%
    {}%
}

\newcommand{\update}[2]{%
  \ifthenelse{\boolean{editing}}%
    {\textcolor{orange}{[Update: \textbf{#1} → #2]}}%
    {#2}%
}

\makeatother

\begin{document}
\settopmatter{printfolios=true}
\title{\systemname{}: Conversation Enhancement via Multi-Earphone Collaboration}

\author{Lixing He}
\email{1155170464@link.cuhk.edu.hk}
\affiliation{%
  \institution{The Chinese University of Hong Kong}
  \city{Hong Kong}
  \country{China}
}
\author{Yunqi Guo}
\email{yunqiguo@cuhk.edu.hk}
\affiliation{%
  \institution{The Chinese University of Hong Kong}
  \city{Hong Kong}
  \country{China}
}
\author{Zhenyu Yan}
\email{zyyan@ie.cuhk.edu.hk}
\affiliation{%
  \institution{The Chinese University of Hong Kong}
  \city{Hong Kong}
  \country{China}
}
\author{Guoliang Xing}
\email{glxing@ie.cuhk.edu.hk}
\affiliation{%
  \institution{The Chinese University of Hong Kong}
  \city{Hong Kong}
  \country{China}
}

\newcommand{\systemname}{CoHear}
\newcommand{\fig}{Fig.~}

\begin{abstract}
In crowded social settings like conferences, background noise, overlapping voices, and lively interactions often lead to ``cocktail party deafness,'' hindering clear conversation. While modern earphones are a promising platform for speech enhancement, existing solutions are limited: they either operate on a single device, ignoring the multi-party nature of conversation, or rely on impractical assumptions like fixed conversation areas and pre-recorded audio. We present \systemname{}, a collaborative system that leverages a network of earphones to holistically model and enhance speech at the conversation level. \systemname{} bridges acoustic sensor networks with deep learning for target speech extraction through two key contributions: 1) a novel, conversation-driven network that dynamically forms groups based on user interaction, using verbal and non-verbal cues (primarily head orientation) for robust, infrastructure-free coordination; and 2) a bandwidth-efficient, robust target speech extraction model that effectively utilizes peer-relayed audio as conditioning signals, even under network constraints. 
\systemname{} is evaluated in both real-world experiments and simulations. Results show that our conversation network obtains more than 90\% accuracy in group formation, improves the speech quality by up to 8.8 dB over state-of-the-art baselines, and demonstrates real-time performance on a mobile device. In a user study with 20 participants, \systemname{} has a much higher score than baseline with good usability. 
\end{abstract}

\keywords{}


\maketitle
\section{Introduction}

\begin{figure}
    \centering
    \includegraphics[width=0.6\linewidth]{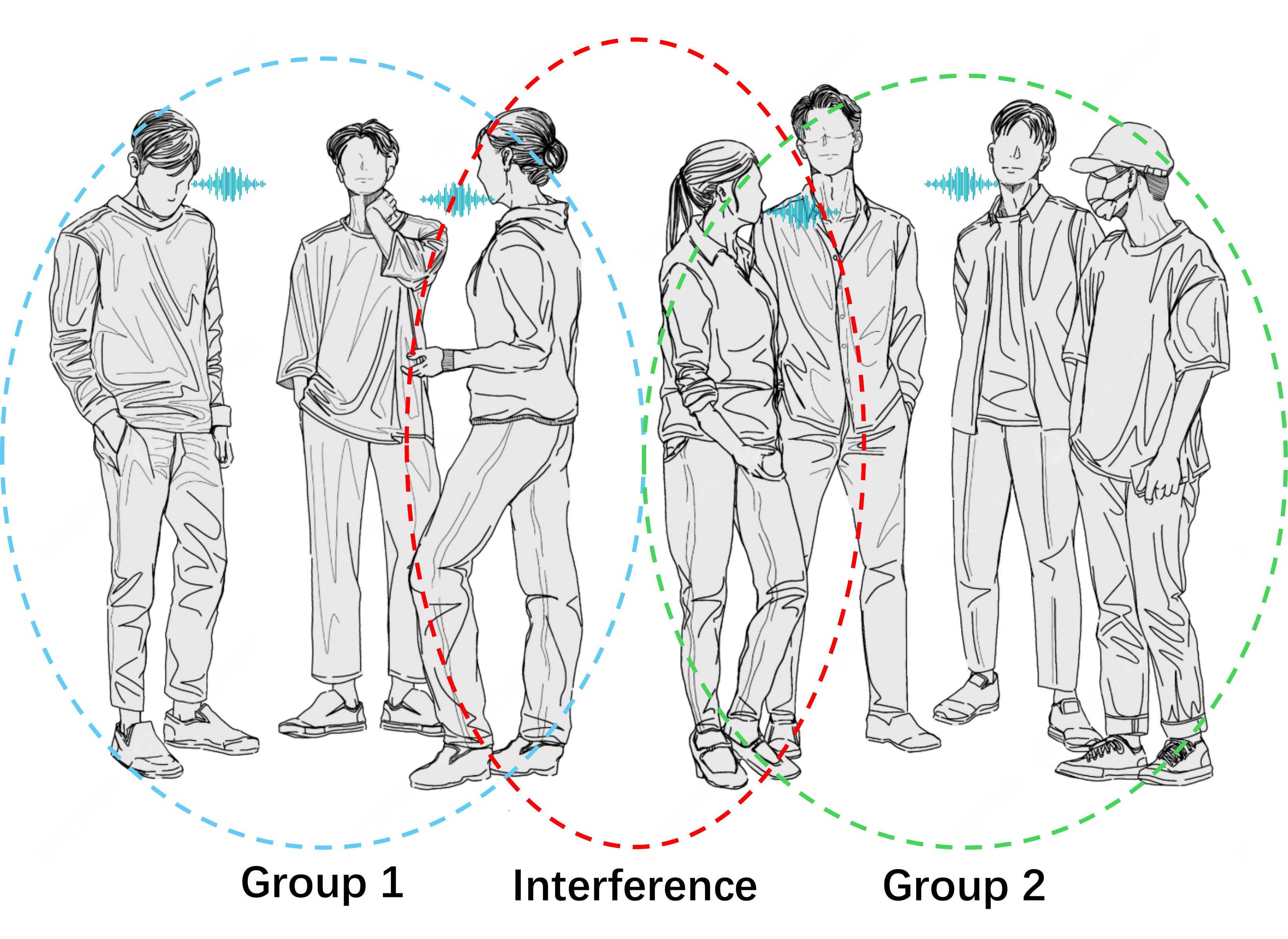}
    \vspace{-1em}
    \caption{Conversation group is naturally formed by interaction with humans, \systemname{} is a collaborative conversation enhancement solution that is driven by the user's intention to talk, automatically find a pal, and deliver clear speech within the same conversation.}
    \label{fig:illustration}
    \vspace{-1em}
\end{figure}

The rising popularity of smart earphones, with approximately 455 million units shipped globally in 2024, marking an 11.2\% year-on-year growth~\cite{patentlyapple2025audio}, frames these devices as a new popular category after smartphones.
In addition to the original usage of playing music, current earphones offer unprecedented convenience and rich functionalities. For example, recent studies on earphones enable powerful speech enhancement~\cite{chatterjee2022clearbuds, he2023towards, duan2024earse}, speech anti-tampering~\cite{duan2024f2key, hearid}, and recognize fine-grained human activities~\cite{lyu2024earda, he2025embodiedsense}. Besides, the head-tracking technology works well to enable spatial audio, which is made possible by motion sensors in earphones \cite{han2023headmon, han2023headsense, lyu2024earda}. Combining the above advantages, we observe that earphones are a promising platform for monitoring human conversation, as they consider both the verbal and non-verbal (head orientation) features.

Conversation is an interactive form of communication between two or more people and plays a crucial role in socialization. As illustrated in Fig.~\ref{fig:illustration}, people naturally form separate conversation groups, and ideally, each person wants to focus solely on the members of their own group, regardless of distance or competing voices. The ``cocktail party effect'' \cite{pollack1957cocktail} refers to the human ability to selectively concentrate on a particular voice in noisy environments. However, this perceptual mechanism has limitations. \new{Many individuals experience what is known as ``cocktail party deafness,'' which hinders their ability to follow conversations in crowded settings due to hearing impairments, cognitive overload, or overwhelming background noise \cite{pryse2009companion, cherry1953some}.} As a result, clear communication in environments such as conferences, networking events, or guided tours becomes increasingly challenging.

To improve hearing during conversations, using earphones as assistive tools presents a promising direction. However, existing works on speech enhancement fall short of our needs, as they do not specifically target conversational contexts. For example, some methods enhance hearing based on sound categories~\cite{veluri2023semantic}, speaker recognition~\cite{veluri2024look}, or listener distance~\cite{chen2024hearable}. Yet, these factors may not accurately capture the dynamics of natural conversation. We cannot assume that all conversations occur within a defined area, which contradicts the assumptions made by SoundBubble~\cite{chen2024hearable}. In contrast, the method proposed in~\cite{chen2024target} is the only one that effectively extracts conversations based on turn-taking dynamics and previously recorded audio. However, it suffers from a major limitation: the reliance on turn-taking events introduces significant delays, rendering it impractical for real-time processing.

To address the limitations outlined above, we have developed \systemname{}, a system that leverages multi-device collaboration from a holistic perspective to enhance the conversation. In contrast to most existing research on earphones, which focuses on single-device capabilities~\cite{dong2024rehearsse, he2023towards, duan2024earse}, \systemname{} acknowledges that conversation is inherently a multi-user activity. By transforming individual earphones into a networked system, our approach integrates the advantages of wearable technology with the strengths of distributed sensor networks.

\new{Our earphone network departs from the conventional assumptions of wireless acoustic sensor networks in two fundamental ways~\cite{gburrek2021geometry}. Traditionally, such networks assume that (1) sound sources appear randomly and independently of sensor node placement, and (2) the sensor nodes themselves remain static. In contrast, our conversation-driven system operates under a reversed premise: the sound sources—human speakers—are co-located with mobile sensor nodes (earphones), and these nodes dynamically organize themselves into conversation groups, as illustrated in Fig.~\ref{fig:illustration}. 
Consequently, the conversation network is triggered and constructed by the interaction between the user, which can be modeled by either verbal or non-verbal features. Among those features, we observe that the verbal features (i.e., turn-taking) are less stable due to the waiting time and prone to interference; we mainly rely on the non-verbal features (head orientation) to determine the group of conversation.

To establish the conversation network, we have access to both multi-channel audio and IMU recordings, which enables the detection of nearby speakers. We envision using these observations to estimate speaker locations and subsequently detect conversations based on their relative orientations. Ultimately, by replaying the enhanced speech through noise-canceling earphones, we aim to augment human auditory perception, effectively outperforming the natural cocktail party effect. However, several challenges are associated with implementing this system:
\begin{itemize}
    \item \textbf{Dynamic peer discovery and network formation.} Since conversations occur spontaneously and participants are mobile, the system must continuously identify nearby speakers, localize their positions, and group devices into socially coherent clusters without any fixed infrastructure.

    \item \textbf{Topology estimation and coordination.} 
    The estimation and observations of a single user may be insufficient to identify conversation participants. Therefore, aggregating the noisy, distributed observations from heterogeneous devices is necessary for establishing a global coordinate system to locate conversations. 

    \item \textbf{Conversation enhancement.} 
    Even a well-established conversation network may not be directly useful to users without additional features. A key enhancement, enabled by the earphone platform, is the targeted extraction of conversation-level speech. The central challenge, which remains unaddressed, is how to effectively select and leverage conversation conditions (e.g., speaker embeddings) from the network to achieve this.
\end{itemize}

To address these challenges, we introduce three key contributions of \systemname{}:
\begin{enumerate}
    \item \textbf{A conversation network for earphones.} 
    We design a novel network that leverages the sensing capabilities of earphones to model conversation groups, forming a mobile, infrastructure-free conversation network.The network includes two key components:
    (i) a node discovery module that detects nearby speakers along with their locations and speaker embeddings, as described in Sec. \ref{sec:acoustic}. This task is divided into two parts: sound source localization and speaker embedding extraction. A joint identity extraction model is used to integrate these two elements.
    (ii) a geometric calibration pipeline that estimates the global network topology using the output from node discovery, as detailed in Sec. \ref{sec:geometric}. This pipeline matches sound sources identified across different devices, calculates the distances between them, and uses this information for geometric calibration.

    \item \textbf{A robust target conversation enhancement.}
    Based on the network structure, we develop a conversation enhancement module that is designed for low latency and high bandwidth efficiency. This module uses peer-relayed audio as conditioning signals and is robust to network delays, packet loss, and compression artifacts. We propose an adaptive model that can utilize different types of conditioning features, including both time-invariant and time-variant features. An audio control module manages the transition between these feature types and regulates the corresponding bandwidth usage.

    \item \textbf{A practical implementation with real-world validation.}
    \new{We implement \systemname{} on earphones equipped with microphones and IMUs, which is practical to be implemented on a commercial device.}
    We evaluate its performance through real-world experiments, simulation data, and user studies. Our system achieves over 90\% accuracy in conversation group formation, improves speech clarity by up to 8.8 dB SNR over state-of-the-art baselines, and demonstrates real-time performance (RTF > 1) on mobile devices. An adaptive bandwidth controller further improves performance under constrained conditions.
    We will open-source our project after publication.
\end{enumerate}}
\section{Related Work}
\label{sec:related}

\subsection{Conversation Modeling}
Conversation involves the exchange of thoughts, feelings, and ideas between two or more people. A natural conversation is not simply a combination of multiple speeches; it also includes both verbal and non-verbal elements. Modeling conversations has significant technical and social implications.

\paragraph{Non-verbal feature}
\new{
Non-verbal features, also known as body language \cite{d2010multimodal}, play a crucial role in conversations. Previous research has explored detecting social interactions using acceleration data from neck-worn sensors \cite{gedik2018detecting}. However, acceleration-based methods are inherently limited to capturing the wearer's own movements and cannot perceive contextual factors or the behaviors of others. Their effectiveness relies on coordinated actions between individuals, which are often inconsistent.

In contrast, vision and audio offer a more direct approach to modeling social interaction. For instance, the Ego4D benchmark defines interaction through the objectives "look at me" and "talk to me" \cite{grauman2022ego4d}. Similarly, multi-modal methods like visual-guided beamforming (e.g., in EasyCom \cite{donley2021easycom}) outperform audio-only approaches by leveraging both data types. However, due to the computational overhead of vision, audio-only solutions remain advantageous. These methods typically work by estimating the speaker's location \cite{wang2022nerc, yang2022deepear}, but this alone is insufficient, as it fails to capture the speaker's orientation—a key component of "talking to me." Some research has aimed to jointly estimate location and orientation \cite{ahuja2020direction}, but these methods are designed for static microphone arrays and lack the adaptability needed for dynamic, real-world environments.
}

\paragraph{Verbal feature}
In addition to non-verbal cues, the content of speech is also a vital aspect of conversation, as highlighted in \cite{sacks1974simplest}. A closely related task is speaker diarization \cite{bredin2023pyannote}, which identifies "who spoke when" and can further transcribe the corresponding text. By analyzing the conversation content \cite{wei2022conversational}, we can generate meeting summaries \cite{ramprasad2024analyzing} or improve readability \cite{wang2024diarizationlm}. Beyond semantic analysis, the dynamics of turn-taking in conversations are noteworthy, as they exhibit clear temporal patterns and are essential for understanding social interactions \cite{levinson2015timing}. 
Previous studies have attempted to utilize the turn-taking to either proactively mediate the conversation \cite{lee2018flower} or detect the turn-taking events by examining natural overlaps and backchannels \cite{chen2024target, chang2022turn}. However, it is important to note that turn-taking detection is only effective once it has occurred, which means that waiting for enough data can pose a significant delay. 

\new{
In summary, non-verbal features are more stable for modeling conversation and introduce less delay than verbal features, making them more suitable for deployment in our system. However, current methods fail to extract these features without visual data, which is unavailable for standard headphones.
}

\subsection{Acoustic Sensing Systems}
\new{Recent advancements in mobile computing leverage acoustic modalities to enable human-centric applications, laying the groundwork for \systemname{}. These works can be categorized into two primary areas: single device and multiple devices, by the number of devices utilized in the system.}

\new{
\paragraph{Single device}
The acoustic sensing capabilities of a single device can be further categorized based on the frequency band utilized. Specifically, when only audible sound (below 20 kHz) is employed, this is referred to as audible acoustic sensing. Conversely, when the ultrasonic frequency band is used, it is classified as ultrasonic acoustic sensing.
}
Audible acoustic sensing extracts semantic content, such as speech transcription \cite{gao2023funasr} or sound classification \cite{gemmeke2017audio}, or derives spatial information through multi-channel audio recordings \cite{wang2019doa}. However, its applications are constrained to audible sources, making it less versatile in complex environments. In contrast, ultrasonic sensing uses controlled, inaudible sound waves to convey information, supporting a diverse range of tasks: underwater localization and communication \cite{chen2022underwater, chen2023underwater}, backscatter \cite{jang2019underwater}, air-to-water communication \cite{tonolini2018networking}, SOS signal transmission \cite{yang2023aquahelper} and dual-channel communication \cite{qian2021aircode}. Except for transmitting information, it can be employed to sense different features, including gesture recognition \cite{wang2020push}, speech enhancement with anti-counterfeiting watermark~\cite{sun2021ultrase, duan2024earse, duan2024f2key}, vital sign monitoring \cite{wang2022loear}, and enhancement with metasurface \cite{zhang2023acoustic}.

\new{
\paragraph{Multiple devices}
Incorporating multiple acoustic devices, the collaboration of them naturally transforms the problem into a network problem, which is called Wireless Acoustic Sensor Network (WASN). Using ambient sound, WASN can determine sensor locations through geometric calibration \cite{gburrek2021geometry}.
}
This process involves iteratively solving a cost function based on the observations of random sound sources. Specifically, geometric calibration can be categorized based on the type of output from the sensor node, including DoA \cite{wang2019doa}, distance \cite{gburrek2021iterative}, or both \cite{gburrek2021geometry}). Instead, there are various algorithms, like iterative optimization \cite{wang2019doa, fischler1981random}, and dataset matching \cite{gburrek2021geometry, hennecke2009hierarchical}).
In addition to geometric calibration, other promising areas to explore include sample rate offset (SRO) \cite{gburrek2022synchronization}, acoustic beamforming \cite{gburrek2023integration}, and swarm-based acoustic network \cite{itani2023creating}.

\new{
In summary, multi-device collaboration enables fine-grained modeling and richer information, which is important for acoustic sensing systems. However, there is a gap in applying these techniques to conversation modeling and enhancement.
}

\subsection{Audio Enhancement}
Prior research on audio augmentation in wearable devices provides a foundation for enhancing conversations in noisy environments, as targeted by \systemname{}. Combined with active noise cancellation, we can improve the user's listening experience seamlessly if the latency is as low as 20-30 ms \cite{stone1999tolerable, gupta2020acoustic}. 
\new{We categorize relevant work into three areas: audio augmented reality, speech enhancement, and non-speech enhancement.}

\paragraph{Audio augmented reality}
Location-based audio enhancement delivers context-specific audio based on a user’s position or orientation. For instance, museum audio guides, such as those described in \cite{harma2004augmented}, play narrations triggered by a user’s proximity to exhibits. Ear-AR \cite{yang2020ear} enhances immersion by aligning stereo audio with the user’s head-related transfer function, using head-tracking to create spatial soundscapes. 
While effective for static, single-user experiences, these approaches assume predefined audio triggers and do not address dynamic, multi-user interactions like conversations, where relative user positions and intentions (e.g., head orientation toward a speaker) are critical.

\paragraph{Speech enhancement}
\new{
Speech enhancement extracts speech from a noisy mixture, either by assuming a model for the noise distribution \cite{wiener1949extrapolation} or by leveraging conditions correlated with the speech signal itself.
These conditions can be derived from wearable devices, such as bone-conducted vibrations \cite{he2023towards}, the in-ear occlusion effect \cite{han2024earspeech}, and facial acoustics \cite{zhang2025wearse}. The data from these sensors is closely correlated with the user's voice, making these approaches user-centric and often designed with a strong emphasis on computational efficiency.
Alternatively, voiceprints (i.e., speaker embeddings) can also be used for speech enhancement \cite{zmolikova2023neural, veluri2024look}. These can be obtained through an audio-only enrollment process.
Systems like SoundBubble \cite{chen2024hearable} can enhance conversations around a user. However, this approach constrains participants to a fixed, circular area, which limits its applicability in dynamic scenarios.
}

\paragraph{Non-speech enhancement}
Other approaches filter non-speech sounds by type \cite{veluri2023semantic}, embedding of the sound \cite{liu22w_interspeech}, or use prior ambient audio for noise cancellation \cite{shen2018mute}. 
However, these methods often focus on single-user scenarios or specific sound categories, which means they do not effectively capture the complex dynamics of group conversations, such as turn-taking in verbal communication and head orientation in non-verbal communication. 

\new{
In summary, a rich body of literature covers various ways to process audio and enhance listening. However, there is limited work on conversation-driven audio enhancement, especially for online deployment as opposed to offline processing.
}

\section{Background}
\label{sec:background}

This section lays the technical foundation for \systemname{}, focusing on acoustic sensing techniques that are important for conversation modeling.
First, we discuss sound localization, which accurately identifies speaker positions using earphone microphone arrays, providing essential information to create a conversation map.
Next, we review the Wireless Acoustic Sensor Network (WASN), which facilitates collaborative sensing across multiple devices.

\subsection{Sound Source Localization}
\new{
Sound source localization is one of the hot topics for multi-channel audio, which enables spatial sensing capability for audio.
Consider a microphone array consisting of $N$ microphones located at known positions $\mathbf{p}_i = (x_i, y_i)$ for $i = 1, 2, \ldots, N$. Let the position of the sound source be denoted by $\mathbf{s} = (x_s, y_s)$.
The time it takes for the sound to reach the microphone $i$ can be expressed as:
\(t_{s, i} = \frac{d_{s, i}}{c}, d_{s, i} = \sqrt{(x_s - x_i)^2 + (y_s - y_i)^2}\)
where $d_{s, i}$ is the distance from the sound source $s$ to microphone $i$, and $c$ is the speed of sound in air. Since the sound source is unknown to the receiver, we can not directly estimate $t_{s, i}$ but calculate the time difference of arrival (TDoA) $t_{s, i} - t_{s, j}$, where $i, j \in (1, N)$. To estimate the TDoA from in-the-wild recording, one common approach is the General Correlation Coefficient (GCC) \cite{knapp1976generalized}, which is computed for each pair of microphones. By finding the peak of GCC, we can estimate the TDoA and calculate the direction of arrival (DoA) by \(DoA = arctan((x_s - x_c)/(y_s - y_c))\), where $x_c$ refers to the center of the microphone array. As shown in the above analysis, the sound localization's performance is heavily reliant on the number of microphones (more microphones, more pairs of microphones, so more TDoA). 

Instead of relying on analytical methods, sound source localization can be performed using a deep neural network, as explored in \cite{wang2022nerc}. This deep learning-based approach generally consists of two stages: feature extraction and location estimation.
In the feature extraction stage, the audio signal is typically converted into a spectrogram representation with dimensions: \(C\times T \times F\), where \(C\) refers to channels, \(T\) refers to time, and \(F\) refers to frequency. These features can be further processed using techniques such as log-melspectrogram conversion, which is applied to each channel individually, or multi-channel methods like GCC \cite{knapp1976generalized} and SALSA-Lite \cite{nguyen2022salsa}.
In the location estimation stage, the neural network is trained to output the sound source location, which also supports a dynamic number of simultaneous sound sources. For instance, architectures employing multi-track \cite{cao2021improved} or multi-ACCDOA \cite{shimada2022multi} output formats are designed specifically to localize overlapping sound events.
}

Besides, deep sound localization can extract the rich information embedded in binaural recording (microphones on the ears), which comprises the complex reflection from the human head and body. Specifically, the head-related transfer function (HRTF) can be modeled as:
\(H_{L}(\theta, \phi, f, d) = \frac{Y_{L}(f)}{X(f)}, H_{R}(\theta, \phi, f, d) = \frac{Y_{R}(f)}{X(f)},\) where $H_{L}$ indicates the frequency response for left ear and $H_{R}$ indicates the frequency response for right ear. 
By leveraging HRTF and deep learning, we can achieve higher accuracy than using a linear two-microphone array. For instance, this approach can partially resolve front-back confusion, a problem that is particularly challenging for conventional two-microphone arrays.

\subsection{Wireless Acoustic Sensor Network}
We are studying a wireless acoustic sensor network where a group of sensor nodes is randomly distributed in a reverberant environment \cite{bertrand2011applications}. We assume that the internal geometric configuration of each node's microphone array is known and that all microphones within the array are sampled at the same time, which we believe is a reasonable assumption following \cite{gburrek2021geometry}. Additionally, we assume that there is coarse time synchronization among the clocks of the sensor nodes, enabling us to aggregate the data globally. By default, we assume a two-dimensional space, but it can be easily extended to a three-dimensional space.

Since a sensor node cannot determine its own position or orientation within the global coordinate system, all observations are made in the local coordinate system of each node. 
Each sensor node \( l \) (where \( l = 1, \ldots, L \)) computes acoustic features (e.g., DoA estimates \( \phi_k^l \) and distance \( d_k^l \)) to the acoustic source \( k \), all reference to the node's local coordinate system. Consequently, we gather a total of \( K \times L \) estimates as the inputs.

From the global view of the WASN, it is composed of \( L \) sensor nodes each equipped with a microphone array located at positions \( n_l = [n_{l,x}, n_{l,y}]^T \) and oriented at angles \( \theta_l \) (where \( l = 1, 2, \ldots, L \)) relative to a global coordinate system defined by the x and y axes. There are \( K \) acoustic sources located at positions \( s_k = [s_{k,x}, s_{k,y}]^T \) (where \( k = 1, 2, \ldots, K \)). Note that all the above global notations are unknown and will be estimated through a geometry calibration process based on the observed acoustic signals from sensor nodes.

\section{System Design}
\label{sec:system}

\subsection{Design Overview}
\label{sec:overview}
\begin{figure}[h]
    \centering
    \includegraphics[width=1\linewidth]{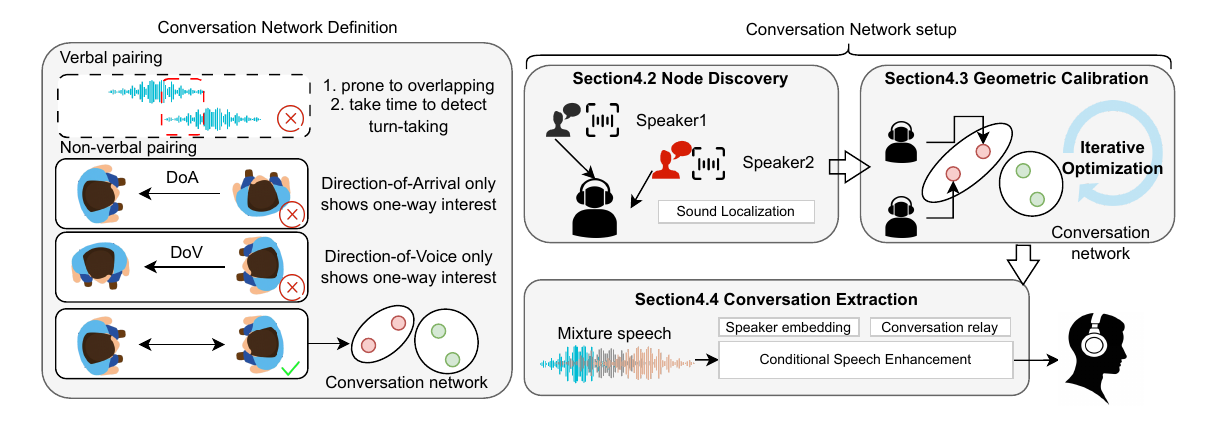}
    \vspace{-1em}
    \caption{Overview of the \systemname{} system. We conduct node discovery for each user and carry out geometric calibration collaboratively. Finally, we perform target conversation extraction based on clues related to the members of the conversation.}
    \label{fig:system overview}
    \vspace{-1em}
\end{figure}

\new{
\systemname{} is a novel network system specifically designed to establish connections based on conversational features, as illustrated in Fig. \ref{fig:system overview}. To achieve this, we define a "conversation network" as the foundation of the system, where individuals participating in the same conversation will experience conversation enhancement.
To construct such a network, the first step is to define its basic components: pairings and which participants are considered part of the same conversation. The details of this definition will be discussed in the next subsection.
Given the definition of the conversation network, the next step is to implement it in the real world using earphones, as shown on the right side of Fig. \ref{fig:system overview}. \systemname{} employs a two-stage process for conversation modeling. In the first stage, we set up the conversation through acoustic sensing and optimization, and in the second stage, we perform enhancement powered by the network.

The implementation consists of three core components:
First, the \textbf{node discovery} module acts as the front end of our system, where nearby nodes (i.e., speakers) are identified using deep learning-based sound source localization. We extract both the locations and embeddings of each speaker to serve as input for the next stage. Note that this process applies to all speakers wearing earphones; further details are provided in Sec. \ref{sec:acoustic}.
Second, \textbf{geometric calibration} is performed in a centralized manner by aggregating the outputs of the node discovery module from each speaker, as discussed in Sec. \ref{sec:geometric}. Although each node can only observe others within its local coordinate system, we iteratively optimize a shared global geometry. The more observations we obtain from the speakers, the more stable the optimization becomes. After completing the initial optimization with a cold start, we can track the speakers in real time.
Once the conversation network is established, the final component, \textbf{conversation enhancement}, performs conversation-level speech enhancement based on the identified participants (described in Sec. \ref{sec:conversation}). Specifically, both the speaker embedding (time-invariant) and the relay signal (time-variant) are used as cues for the conditional speech enhancement model. We introduce an audio control module to balance the contributions of these two approaches.
}

\paragraph{Conversational network}
To define our conversation network, the first step is answering the question: How can we technically model the conversation?'' Specifically, we break down a conversation into individual interactions performed by each participant, such as when user A speaks to user B. We observe that such a definition is similar to the concept of computer networking, where a sender sends pairing messages to find a receiver.
Accordingly, we construct a conversation network comprising all participants that can be successfully paired with the same conversation.
It is important to note that existing platforms, such as Livekit \cite{livekit}, provide real-time audio-visual sharing via WebRTC \cite{blum2021webrtc} Selective Forwarding Units (SFUs). In contrast, \systemname{} builds upon these platforms and extends their capabilities with advanced conversation modeling.

\new{
To enable the pairing of networks, we consider two features as discussed in Sec. \ref{sec:related}: verbal and non-verbal. Since the verbal feature (turn-taking) needs to wait for enough time before being effective, we consider the non-verbal feature as the main factor to determine the conversation.
We consider that a person's head orientation is a key indicator of their willingness to engage in conversation, both to speak and to listen. We define a conversation as requiring a bidirectional interest: individuals in the same conversational group must have an interest in at least one other person within the group. This mutual interest is depicted in the bottom left of Fig. \ref{fig:system overview}. As this pairing relies solely on non-verbal cues, we refer to it as the non-verbal pairing of a conversation network.
If we aim to implement this pairing using audio, sound source localization is a natural consideration. However, while sound source localization can estimate a person's relative position, it is insufficient for identifying conversational groups. This is because the DoA of sound can only determine a unidirectional interest (i.e., who is being listened to). Similarly, Direction-of-Voice also only indicates the other way of interest, so that both DoA and DoV are not the indicators for conversation.
Formally, combining both DoA and DoV allows us to infer the desired 6 Degrees of Freedom (6DoF) locations of participants. This necessity motivates us to move beyond the limitations of existing sound source localization methods.

Verbal features, such as turn-taking, serve as direct indicators of conversational engagement, as illustrated in the upper-left section of Fig. \ref{fig:system overview}. However, a significant drawback is their reliance on observing a sufficient duration of speech to be effective (i.e., one must wait for a turn-taking event to occur). Furthermore, overlapping audio is prone to interference from other speakers, which reduces reliability.
Consequently, the conversation network in our current implementation is primarily constructed using non-verbal pairing. Although these non-verbal features are also extracted from the audio signal—the same source as the verbal features—they are processed in a fundamentally different way. This alternative processing method ensures that the two aforementioned disadvantages do not apply to our design.
}

\begin{figure}[h]
    \centering
    \includegraphics[width=0.7\linewidth]{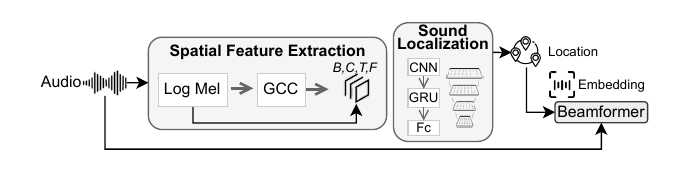}
    \vspace{-1em}
    \caption{The node discovery for conversation setup: where we detect the sound source in the first stage and find the corresponding speaker embedding in the second stage.}
    \label{fig:acoustic}
    \vspace{-1em}
\end{figure}

\subsection{Node Discovery for Conversation Setup}
\label{sec:acoustic}
Conversation can be depicted through the interaction between speakers; the speaker presents the interest according to the direction of the voice (also the direction of the head). In \systemname{}, each user is considered as one node, where we reformulate speaker localization as a dynamic node discovery problem in a mobile acoustic network, as shown in Fig. \ref{fig:acoustic}. Specifically, it is necessary to access both the location and identity of the nearby nodes. We represent the location in local coordinates (the receiver as the root) and the identity as the voiceprint, which is an embedding that represents the voice of the speaker.

While joint estimation of these properties (similar to sound event localization and detection) is theoretically possible, prior work shows such approaches suffer from computational complexity and poor scalability \cite{wang2022nerc}. Instead, we adopt a decoupled two-stage discovery process that separately handles node localization and identity extraction and then fuses the results for network coordination.

\paragraph{Stage one: sound localization}
\new{
As for the sound localization, we extract the cross-channel features as Sec. \ref{sec:background} and utilize a convolutional recurrent neural network (CRNN) to process them, which consists of multiple convolutional layers, a GRU layer, and the final linear layer to predict the output. 
Specifically, we convert the multi-channel audio recording ($M \times T$) into multiple pairs of recordings ($N \times 2 \times T$), where $N$ refers to all the possible combinations. Then, we apply a log-mel spectrogram transformation for each channel and calculate GCC between every pair of channels. Lastly, we concatenate every pair of channels with the original log-Mel spectrograms as the final representations. For example, the feature has a shape of $(2+1) \times T \times F$ for two-channel audio recordings, which can be considered as the input to the neural network. As for the output format, we represent the sound source location in a cartesian coordinate system ($x, y, z$) with a normalized range of one meter radius ($x^2+y^2+z^2 = 1$), where the output dimension of the last linear layer is 3 with a tanh activation. When there is no sound source, the expected $x,y,z$ are set to zero. During inference, we only consider $x^2+y^2+z^2 > 0.5$ as a valid sound source.
Considering the output frame length is 0.1 seconds, we need to set the downsample ratio of the neural network to match it. Given the hop length of the spectrogram is 20ms, the time-dimension downsampling is set to 5. Consequently, the output format is $3 \times T/5$.
}

The above design only fits in one-source localization; we incorporate the multi-track \cite{cao2021improved} format for multi-source sound localization. Since we are only interested in the location rather than the sound class, the multi-ACCDOA \cite{shimada2022multi} format is unnecessary.
Considering the max number of sources is $n_{max}$, the output of the model is adjusted to $3*n_{max} \times T/5$ instead. Similar to sound separation, multi-source sound localization also needs permutation-invariant training as follows:
\(loss = \min_{\pi \in \Pi} (MSE(pred_\pi, gt))
\), where $\Pi$ refers to all the possible permutations of the sound sources, the above loss intends to find the optimal permutation (the least loss) and then perform back-propagation.

\paragraph{Stage two: identity extraction}
\new{
After localizing all the sound sources, the next step is to associate the sound location with the identities. Although the speaker identities are accessible with the help of existing tools like Pyannotate \cite{bredin2023pyannote}, which supports overlapping speech and multiple speakers. 
However, the association between locations and identities can be difficult since they are two independent attributes. Specifically, the location of each person can be permuted randomly without impacting the results.

To associate them effectively, we can perform beamforming for each detected sound source and check which speaker's sound is amplified, which brings two subsequent problems: 1) how to perform beamforming with a limited microphone array, and 2) how to determine the identity of the beamformed speech. Although there are existing solutions for them respectively, directly combining them can lead to degradation since they are not co-optimized. 
For example, the output from beamforming can still contain noise, so that speaker identity extraction may not handle it well.
}

We propose a joint model to solve both the beamforming and embedding extraction at the same time. Specifically, we keep the beamformer (e.g., delay-and-sum beamformer) the same and reshape the output from \(B \times C \times T \times F\) to \(B \times T \times (F \times C)\) and use a linear projector with layer normalization to predict the d-vector. In addition, the final d-vector is the average over time. Since the output is not audio, we can not use the common loss function for neural beamformer (i.e., scale-invariant signal-noise ratio), but the cosine similarity loss as follows:
\(\text{L}(\mathbf{A}, \mathbf{B}) = 1- \frac{\mathbf{A} \cdot \mathbf{B}}{\|\mathbf{A}\| \|\mathbf{B}\|}\),
where \textit{A} is an estimated d-vector, and \textit{B} refers to the d-vector from clean speech with the speaker verification model.

In summary, the discovered node can be classified into three groups: self-voice, interested voice, and other voices. Note that the self-voice and interested voices come from a similar direction (i.e., the front of the user) with different distances, so it is hard to differentiate them with direction information only. Instead, we assume that the self-identity is given as prior knowledge, which is practical through offline or online enrollment. As a result, the self-voice can be identified by the similarity with the stored speaker embedding, so that we will not discover it as a new node.

\begin{figure}[h]
    \centering
    \includegraphics[width=0.7\linewidth]{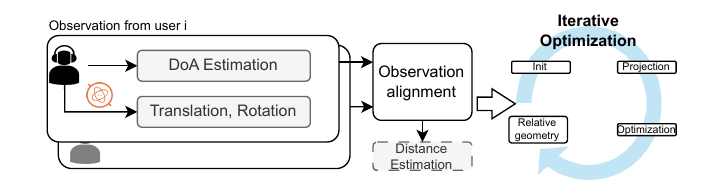}
    \vspace{-1em}
    \caption{Network geometric calibration uses the observed distance and direction of arrival, along with user motion, to estimate the locations of all users in the network via iterative optimization.}
    \label{fig:geometric_calibration}
    \vspace{-1em}
\end{figure}

\subsection{Network Geometric Calibration}
\label{sec:geometric}
As outlined in our system overview (Sec. \ref{sec:overview}), our ultimate goal is to estimate the positions of all users (nodes) in a conversation network. The previous node discovery step identifies nearby speakers, providing important positional observations. However, these outputs only indicate where people are speaking, not the structure of the conversation network itself. To achieve this, we propose aggregating observations from multiple participants to estimate the network's underlying geometry.
This task can be formulated as a multi-sensor localization problem, known as geometric calibration. Given that we use acoustic signals for localization, our approach falls within the category of WASN, as discussed in Section \ref{sec:background}.

This section details our geometric calibration pipeline. First, we describe how we obtain the necessary inputs: observation sharing and distance estimation, which provide DoA and distance measurements, respectively. Next, we introduce the concept of a Mobile WASN, distinguishing it from conventional Wireless Acoustic Sensor Networks. We then present our improved geometric calibration algorithm, which estimates the final positions of all nodes. Finally, we address two critical implementation factors: 1) reducing the warm-up time through dynamic calibration, and 2) enabling the system to operate with only a single device.

\paragraph{Observation alignment}
\label{sec:sharing}
The principle of geometric calibration is that one source can be observed by multiple nodes, where the relative position is related to the nodes' positions. Thus, it is important to align the source across nodes.
It is a trivial task when there is only one sound source, but it becomes complicated with 1) overlapping sound, 2) the sound isn't observed by some nodes. Since we have already extracted the corresponding speaker embedding for each source, we can assign the same identity to the sources with similar embeddings.
Specifically, for the speakers observed by one user (including itself), we pair all of them with the observations from other users and only keep the ones that have high similarity. The similarity is calculated by the cosine similarity of two embeddings.
A group of sources with more than 0.8 similarity is considered the same source. We don't make the assumption that one source can be observed by all users.
In practice, every device uploads its observed embeddings from Sec. \ref{sec:acoustic} to a central server and performs the above matching. Only the source that is observed by at least two nodes are considered as valid ones in the following.

\paragraph{Distance estimation}
\label{sec:distance}
According to the output of Sec. \ref{sec:acoustic}, we can only obtain the direction of arrival, while the distance to the sound source is skipped.
However, the distance information is important in geometric calibration since the optimization can be under-determined without it. Ideally, audio recordings already embed distance information in a reverberant environment, as illustrated in the following:
\[
y(t) = h(t) * x(t) + v(t) = h_{e}(t) * x(t) + h_{l}(t) * x(t) + v(t)
\], where \( v(t) \) corresponds to white sensor noise and \( h(t) \) represents the room impulse response, the symbol \( * \) denotes convolution. However, the performance highly relies on a known environment, so the scalability is not satisfying \cite{krause2021joint, gburrek2021geometry}.

\new{
Instead, we propose estimating distance by comparing the sound pressure levels at the transmitter and receiver. In an ideal scenario, the audio attenuation corresponds to the propagation path loss, which is related to distance. To ensure a clean input signal and mitigate the effects of multi-path propagation or overlapping sound sources, we selectively use audio segments containing only a single active source and compute the average sound level from these segments.
Despite this design, distance estimation errors may still be significant. Therefore, we do not use the estimated distance as a final output. Instead, we utilize it as an input for subsequent geometric calibration, alongside other data.
}

\paragraph{Mobile WASN}
According to Sec. \ref{sec:background}, our network can be considered as mobile WASN, an extension of WASN where the acoustic node can move freely. Similarly, localization of sensor nodes can be achieved by geometric calibration, given sufficient observed sound sources. Compared to the conventional notation, mobile WASN has different assumptions as follows.

\begin{enumerate}
    \item First, the assumption that only one sound source is present during a measurement is unrealistic due to the possibility of overlapping speech and other background sounds. Specifically, if we denote the maximum number of simultaneous speech sources as \(M\), we can estimate a total of \(M \times K \times L\) instances, which is also compatible with fewer than \(M\) number of sources by adding a blank estimate.
    \item Second, while all the estimates are considered valid in the previous context, the arbitrary movement of users can create a much greater distance between two sensor nodes, which means that some sound sources may not be observed by all sensor nodes.
    \item Third, the acoustic source \(S_{k}\) is defined as the \(k^{th}\) observed source during a specific time period. In this context, \(k\) (where \(k = 1, 2, \ldots, K\)) can also serve as a time identifier. The notation is important because we can align the sound source with the location of the sensor nodes, which are also time-dependent and can be represented as \(n_{k,l}\). Since the sensor nodes can move freely, we use \(T_{k,l}\) and \(R_{k,l}\) to indicate the translation and rotation between \(n_{k,l}\) and \(n_{0,l}\) for the \(l^{th}\) node.
    \item Lastly, it is expected that \(n_{k,l}\) will match at least one of the \(S_{k}\) since the sensor node is one of the sound sources. Note that it is not valid conversely since there are sound sources that come from people without earphones. 
\end{enumerate}

\paragraph{Geometric calibration}
We perform mobile geometric calibration with the networked earphones, which share a similar principle as other geometric calibrations that minimize a cost that is defined to evaluate how well these converted observations align with the assumed geometry. Among the various algorithms, data set matching is an efficient algorithm for geometric calibration. Our algorithm is based on that of \cite{gburrek2021geometry}; the details of it are out of the scope of this paper.. 

Different from the original algorithm, we have a rotation matrix \(R_{k,l}\) and a translation vector \(n_{k,l}\) that are dependent on both the source and the node. As a result, we rewrite the relative locations in the global coordinates:
\[
G_{k,l} = R_{k,l} d_{k,l} [cos(\theta_{k,l}), sin(\theta_{k,l})] + n_{k,l}
\]
Multiple nodes observe each source at the same time; if all the estimates are perfectly right, all \( S_{k,l}\) would map to a unique position \( G_{k,l} \). Hence, the geometry can be inferred by minimizing the deviation of the projected source positions by the cost function below:
\[
\arg\min \sum_{l=1}^{L} \sum_{k=1}^{K} \left\| G_{k,l} - S_{k,l} \right\|_2^2
\]
, where \( \| \cdot \|_2 \) denoting the Euclidean norm. There exists no closed-form solution for the above nonlinear optimization problem, so it has to be solved using an iterative optimization algorithm.

Compared to our setting, previous methods estimate 
\(R_{l}\) and translation \(n_{l}\) for L sensor nodes, while ours estimate \(K \times L\) nodes equivalently. The naive approach is performing geometric calibration for each source, downgrading to K times one source geometric calibration, and merging the results afterward.
According to the analysis in \cite{gburrek2021geometry}, the performance of geometric calibration depends on the number of sources. Consequently, the one-source geometric calibration will give unreliable results and harm the final performance.

Instead, we can further compose the translation into initial and movement: 
\(R_{k,l} = \overline{R}_{k,l}R_{0,l}\) and \(n_{k,l} = \overline{n}_{k,l} + n_{0,l}\), where \(\overline{R}_{k,l}\) refers the rotation from the first source to the k source for sensor l and \(\overline{n}_{k,l}\) refers the corresponding translation. We can see that if the \(\overline{R}_{k,l}\) is an identity matrix and \(\overline{n}_{k,l}\) is zero, the problem downgrades to the same as \cite{gburrek2021geometry}. As a result, we rewrite the above equation:
\[
G_{k,l} = R_{0,l} d_{k,l} \overline{R}_{k,l} [cos(\theta_{k,l}), sin(\theta_{k,l})] + \overline{n}_{k,l} +  n_{0,l}
\]
Fortunately, we can estimate \(\overline{R}_{k,l}\) and \(\overline{n}_{k,l}\) by a motion sensor on the node, which means our estimation targets become \(R_{0,l}\) and \(n_{0,l}\) so that the problem becomes estimating one locations instead of track. In other words, we can apply the data set match algorithm in the appendix.

In addition, the above processing can be generalized to multiple sources. Suppose \(M\) is the maximum number of simultaneous sources, the \(\overline{K}\) refers to the real number of sources within K times (\(K\leq\overline{K}\leq M \times K\)). We can set the \(R\) and \(n\) accordingly to fit in the original setting.

\paragraph{Dynamic calibration}
Based on our design, we conduct geometric calibration for every \(\overline{K}\) observation, which is not ideal for real-time inference. As a solution, we implement a sliding-window approach that allows us to repeat geometric calibration each time we receive new observations. Given that the runtime latency for this process is minimal, we plan to address the optimization in future work.

Meanwhile, selecting an optimal \(\overline{K}\) becomes critical, as it seems to be related to the latency and performance. Although the correlation between latency is obvious, the correlation with performance can be complicated. Specifically, the motion estimation (\(\overline{R}_{k,l}\) and \(\overline{n}_{k,l}\)) by IMU is also poisoned by accumulated errors with increasing \(\overline{K}\), which means the calibration may not be accurate with a large \(\overline{K}\). A small sliding window can be a direct solution to mitigate the degradation caused by noisy estimates. However, since the optimization needs sufficient observations to converge, the small window can conversely degrade the usability. Instead, we propose a weight decay algorithm where the calibration results focus more on the nearest observations. Specifically, we set the weight as \(  1/ (T-t)\), where \(T-t\) refers to the difference between the current time and the time of observations.

\paragraph{Single-device solution}
\new{
While collaborative calibration is the optimal approach, we propose a single-device method for scenarios where networking is impractical or latency is critical. In the evaluation, the above design will be considered as a special case when there is one earphone.
In geometric calibration, the output for each user is their position (\(P=(x_i, y_i)\))and orientation (\(O=\theta\)) within a shared coordinate system. However, in a single-user scenario, the shared coordinate system is defined relative to the user themself. In this case, the source's position and orientation relative to the user can be described by three key variables: the Direction of Arrival, the distance to the source, and the source's own orientation.

Assuming the sound source localization method from Section \ref{sec:acoustic} provides a reliable DOA, we note that the distance and source orientation variables are still missing. To address this, we modify our model to estimate these two additional variables.
Our neural network already outputs a 3-dimensional vector. We remove the unit vector constraint (\(x^2+y^2+z^2=1\)) to allow the model to represent sources at any distance. Furthermore, we add four additional output nodes to classify the sound source's orientation into one of four categories: front, left, right, or back. We use classification instead of regression for the orientation, as this challenging problem benefits from the increased stability that a classification approach provides.
}

\begin{figure}[h]
    \centering
    \includegraphics[width=0.7\linewidth]{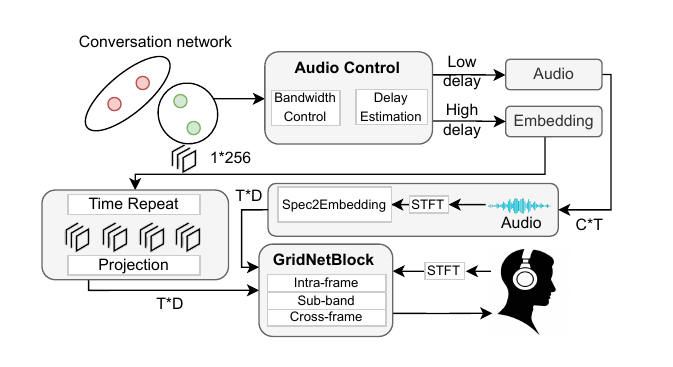}
    \vspace{-1em}
    \caption{Conversation extraction by target feature, which can be classified into time-invariant (speaker embedding) and time-variant. The adaptive audio control can balance the quality and bandwidth before sending to the neural network model.}
    \label{fig:voice_share}
    \vspace{-1em}
\end{figure}

\subsection{Conversation Extraction}
\label{sec:conversation}

\paragraph{Problem formulation}
For each group of nodes participating in the same conversation, the locations, identities (represented by speaker embeddings), and the audio recordings are shared among all members. We utilize this information to adjust the user's hearing (extract conversation) as demonstrated in Fig. \ref{fig:voice_share}.

Our design goal is to extract a target conversation from a speech mixture under specific conditions. While raw audio provides more fine-grained information than speaker embeddings, it consumes significantly more bandwidth. A naive solution, assuming we have access to a relatively clean audio relay, would be to transmit this audio directly like a video conference. However, this assumption is impractical for two reasons. First, even after noise suppression, we cannot guarantee that the transmitted audio perfectly matches the target speech. Second, the transmission process itself introduces delay, packet loss, and compression artifacts, resulting in a degraded version of the original received audio. Therefore, the core problem we address is how to design a conversation extraction model that can effectively utilize this degraded relay audio.

\paragraph{Adaptive audio control}
\new{
Ideally, relay audio should be transmitted at the highest possible quality. However, there will always be a bandwidth bound, particularly with multiple concurrent users. This is a common problem in networking, addressed by algorithms like TCP congestion control.
Unlike these well-established solutions, \systemname{} also considers the acoustic noise level, where a higher noise level indicates the need for a larger bandwidth (to provide more information as compensation). 
To achieve this, we can perform a grid search to maximize the overall conversation quality for all users under a total bandwidth constraint. For each user, we consider the distance as the indicator for conversation quality instead of directly estimating the metric like SNR. Consequently, the search objective is the sum of bandwidth/ distance for all the users within the conversation, where the total bandwidth is constrained. The reason behind our design is the close correlation between pair-to-pair distance and noise level.

Besides the bandwidth, transmission delay is another factor that needs consideration. Specifically, a delayed audio is almost useless even if he audio quality is good since the time-dimension correlation between the time-variant feature and the target speech diminishes, reducing its effectiveness as a reference. 
To address this, we propose estimating the transmission delay and disabling the time-variant feature when the latency exceeds a certain threshold. 
Specifically, we can utilize the GCC technique between the transmitted audio and the noisy speech, identifying the first peak as the delay.
Note that we don't need to estimate the delay frequently since we assume the delay is relatively stable.
In our empirical test, we find that a delay of 50ms is still tolerable for the time-variant feature and set it as the threshold.
Consequently, under poor network conditions, the system gracefully degrades by relying solely on time-invariant features for target speech extraction.
It is also important to note that the switch between time-invariant and time-variant features impacts bandwidth control. When audio transmission is not necessary (i.e., when the time-variant feature is disabled), the bandwidth requirement can be reduced accordingly.
}

\paragraph{Target conversation extraction}
\new{
Given the above analysis, the time-variant feature can be viewed as an extension of the time-invariant feature. Although the former provides richer information, several factors beyond bandwidth—such as latency, noise, and transmission losses (e.g., packet loss, bit errors)—can degrade the quality of these features.
Building on prior work in target speaker extraction \cite{veluri2024look}, we focus on target conversation extraction as an extension of speech enhancement. Existing target speech extraction methods typically rely on either time-invariant features, such as embeddings \cite{chen2024target, veluri2024look} or time-variant data like bone-conducted vibrations \cite{he2023towards}.
Time-invariant features, while compact and stable, can be viewed as ultra-highly compressed audio, potentially losing fine-grained details compared to time-variant data, like bone-conducted vibrations \cite{he2023towards} or relay audio \cite{shen2018mute}. To address this, \systemname{} is designed to strike a balance between compression and performance, optimizing both efficiency and accuracy.
}

Considering the time-invariant feature like speaker embedding, the enrollment speech (from the target user) is first converted to the speaker embedding by an encoder. To fit the feature in the model, the speaker embedding is projected to the same feature dimension and repeated to align with the time dimension. Lastly, the speaker embedding in the time-frequency dimension is added to the enhancement network. The pipeline is illustrated as follows:
\[
y = TSE(y_{noisy}, [e_{enrollment}, ....]_{T}), \quad e_{enrollment} = Embedder(y_{enrollment})
\], where the $TSE$ refers to the target speech extraction model (i.e., modified TFGridNet) and $Embedder$ refers to the speaker embedding extractor. We consider that $y_{enrollment}$ comes from the target user whose voice is mixed with noise in $y_{noisy}$.
As for the time-variant feature, which also refers to the target speech as the speaker embedding. We can transform the relay audio into a time-frequency feature using an STFT encoder. Similarly, the feature is added to the enhancement model, and the pipeline is illustrated below:
\[
y = TSE(y_{noisy}, [e_{enrollment}, ....]_{T}, E_{relay}), \quad E_{relay} = Embedder(y_{relay})
\]. To maintain compatibility with time-invariant features, we include both time-invariant features and time-variant features during the training by setting the features to zero when it is absent. Note that the actual form of the time-invariant feature is not the raw audio; we can further improve the efficiency by only transmitting the feature, and we leave it for future work. 
Besides, \systemname{} is expected to work well with multiple conditions (participants). A naive solution is to apply the speech filter multiple times, which is computation-intensive. Instead, we propose an attention mechanism that combines multiple conditions into the intermediate feature as follows: 
\(
 X = Attention([X_0, X_1, ...])
\)
, where \(X_i\) refers to the \(i^{th}\) conditions.

\section{Evaluation}


\subsection{Implementation and Evaluation Setup}
\begin{figure}
    \centering
    \includegraphics[width=0.5 \linewidth]{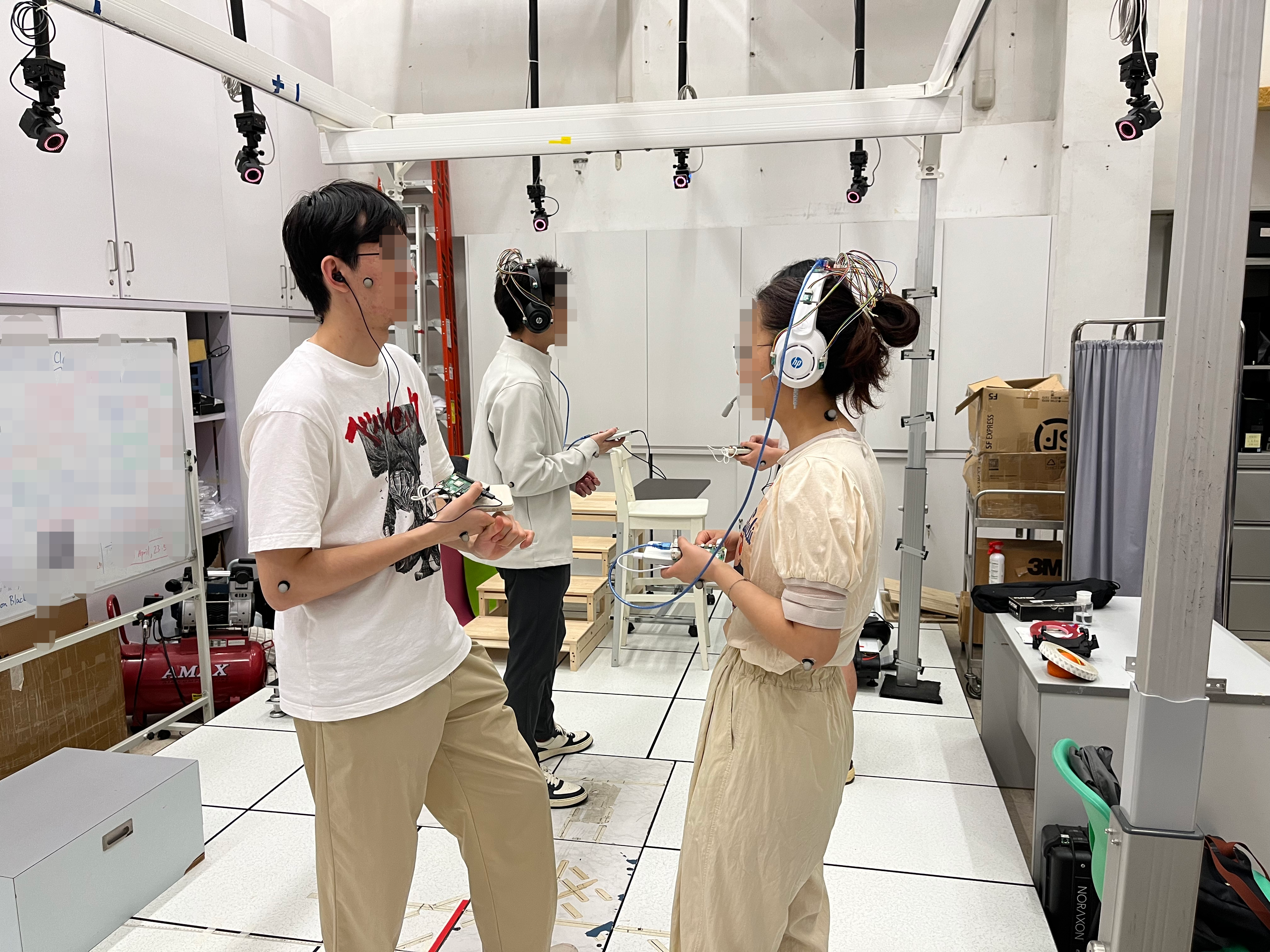}
    \caption{Experiment setup of the motion capture room.}
    \label{fig:mocap}
    \vspace{-1em}
\end{figure}

\begin{figure}[t]
    \centering
    \begin{minipage}{0.48\linewidth}
    \centering
    \includegraphics[width=0.8\linewidth]{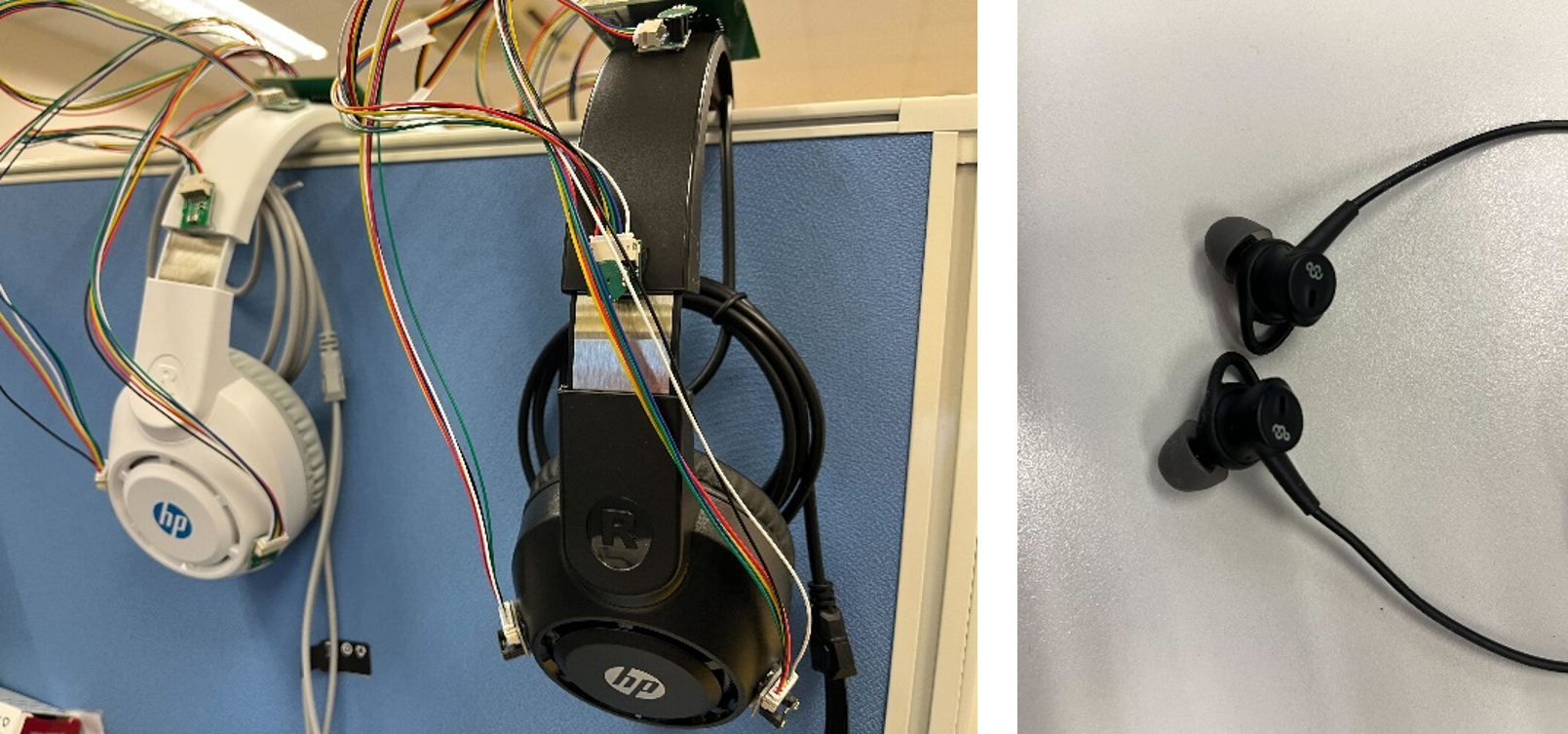}
    \caption{Hardware used in the experiment.}
    \label{fig:hardware}
    \end{minipage}
    \hfill
     \begin{minipage}{0.48\linewidth}
        \centering
        \includegraphics[width=0.4\linewidth]{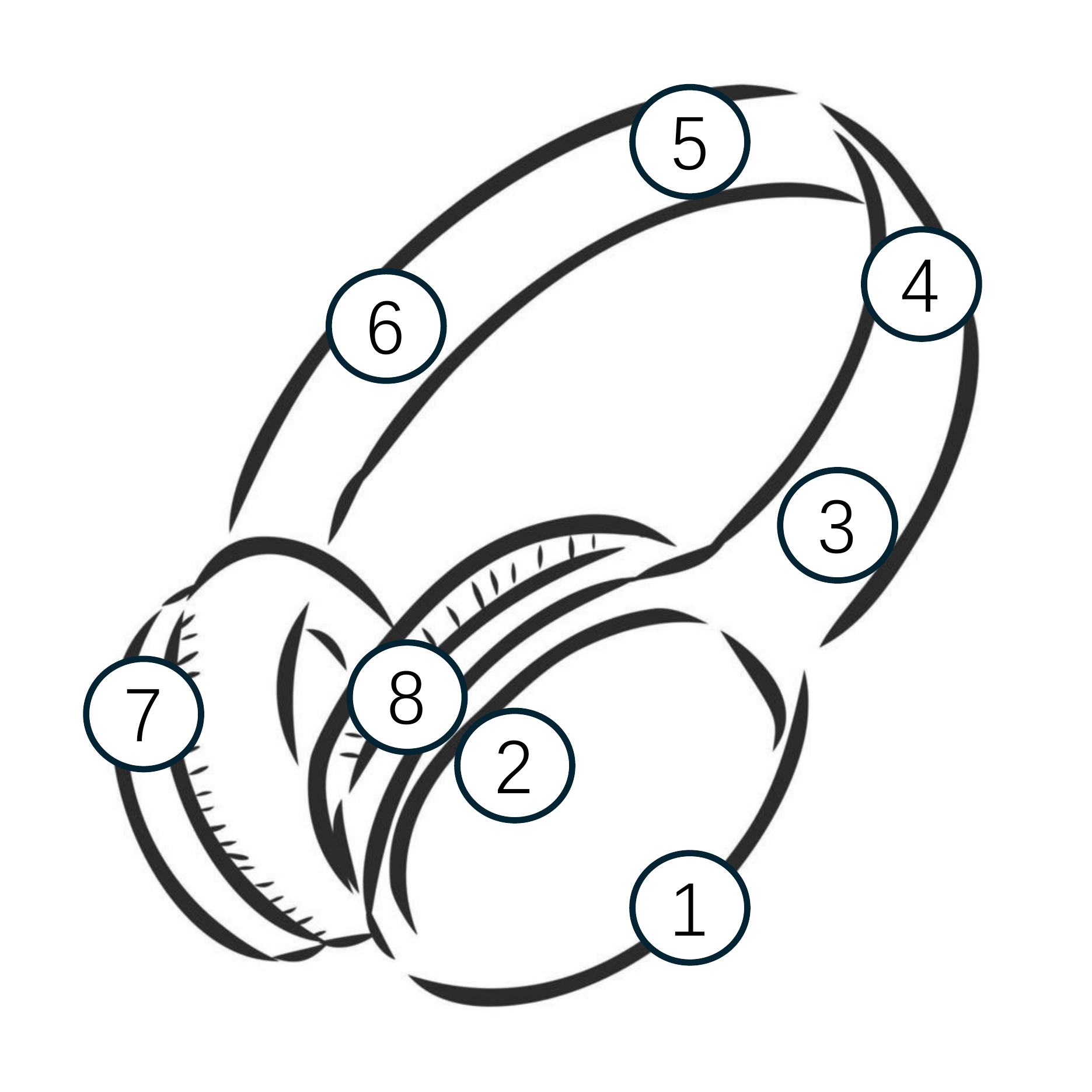}
    \caption{Microphones location on the headphone.}
    \label{fig:mic_loc}
    \end{minipage}%
    \vspace{-1em}
\end{figure}

\paragraph{Testbed Setup}
\label{sec:hardware}
\new{
We implement \systemname{} in various form factors, including binaural earphones and headphones with up to an 8-channel microphone array (placement is illustrated in Fig. \ref{fig:mic_loc}). The two earphones are shown in Fig. \ref{fig:hardware}. We use a Raspberry Pi 5 to collect audio and attach one IMU (BMI-160) to the headphones to track head motion, transmitted via I2C. For experimental validation, we connect each device via SSH under the same WiFi network, streaming all data to a desktop for analysis. The system output is designed to be played back through the headphone speakers, maintaining compatibility with existing ANC functionality. 

While our hardware supports sampling rates up to 48 kHz for maximum capture capability, the system is designed to operate at lower rates suitable for real-world deployment.
For practical speech applications, \systemname{} can operate at 8 kHz with audio compression to minimize bandwidth requirements, which is also evaluated in the later section.
Moreover, \systemname{} supports time-invariant speaker embeddings that require only one-time transmission, significantly reducing data size to 8192 bits in total, which is practical on even constrained networks. We also discuss the networking issue in Sec. \ref{sec:discuss}.
}

\paragraph{Dataset Collection}
We use three data sources to support the design and evaluation of \systemname{}: a real-world dataset collected in a motion capture room, simulated datasets for large-scale testing and training, and public datasets for comparison.

\vspace{1ex}
\noindent\textit{Real-world Collection:}
We collected 30 minutes of multi-person conversation data in a $10 \times 5$ m motion capture room equipped with a 12-camera VICON system (Fig.~\ref{fig:mocap}). Markers were placed on participants’ heads and upper bodies to track conversational poses. Six volunteers wore our prototype headphones (Section~\ref{sec:hardware}) and engaged in natural turn-taking conversations. Due to space limitations, each session involved 2–4 participants in different grouping scenarios:
1) two participants in one group,
2) three participants with one excluded, and
3) four participants split into two groups.

\vspace{1ex}
\noindent\textit{Simulation:}
To scale beyond the limits of physical collection, we simulate two datasets: 1) Geometric Calibration: Using Pyroomacoustics~\cite{scheibler2018pyroomacoustics}, we simulate dynamic conversations in larger rooms (up to 20 m wide) with varied reverberation (RT60 from 0.25 to 0.75).
And 2) Conversation Extraction: We augment the real dataset with noisy-clean pairs and degraded relays (e.g., SNR, compression, bit crush, packet loss) to train and evaluate the enhancement model. The SNR of the noisy speech is set to between -10 dB and 10 dB.

\vspace{1ex}
\noindent\textit{Public Datasets:}
We also compare \systemname{} with existing datasets in Table~\ref{tab:comparison_dataset}. While prior work such as STARSS23~\cite{shimada2024starss23} and 6DoF-SELD~\cite{yasuda20246dof} focuses on sound event localization, and AMI~\cite{kraaij2005ami} and EasyCom~\cite{donley2021easycom} cover conversations, none support collaborative multi-device conversation modeling with 6DoF tracking as \systemname{} does.

\begin{table}[t]
\caption{Comparison with existing datasets.}
\centering
\begin{tabular}{|c|c|c|c|}
\hline
\textbf{Name} & \textbf{6DoF} & \textbf{Multi-device} & \textbf{Target} \\
\hline
STARSS23~\cite{shimada2024starss23} & No & No & Sound events \\
6DoF-SELD~\cite{yasuda20246dof} & Yes & No & Sound events \\
AMI~\cite{kraaij2005ami} & No & No & Conversation \\
EasyCom~\cite{donley2021easycom} & Yes & No & Conversation \\
\textbf{\systemname{}} & \textbf{Yes} & \textbf{Yes} & \textbf{Conversation} \\
\hline
\end{tabular}
\label{tab:comparison_dataset}
\end{table}

\paragraph{Evaluation Metrics}
The primary objectives of \systemname{} are to identify user locations and extract relevant conversations, assessed through distinct metrics.

For geometric calibration, we use two standard metrics: average direction error (in degrees) and location error (in meters). We also incorporate network setup accuracy, which indicates whether the participant can find the correct partner. Specifically, we define \(S_{i,j}\) equals to true if \(|R_{i,j} - O_i| < \frac{\pi}{4}\), where \(R_{i,j}\) is the relative direction from user \(j\) to user \(i\) and \(O_i\) is the orientation of user \(i\). 

As for the conversation enhancement, we will utilize the scale-invariant signal-to-noise ratio (Si-SNR) as a conversation-related metric, noting that the input SNR significantly influences results. Since the average default input SNR is set to 0 dB, allowing the output SNR to reflect improvements over the input.

\paragraph{Baselines Approaches}
Similar to the metrics, we also incorporate different baselines for the geometric calibration and conversation extraction, respectively.
For the former one, we consider \cite{gburrek2021geometry} as the baseline, which is designed for static WASN only. Besides, infrastructure-based acoustic localization \cite{gburrek2021geometry} and acoustic-SLAM \cite{evers2018acoustic} are related to \systemname{}. Note that when we apply \cite{gburrek2021geometry} as infrastructure-based acoustic localization, we have static nodes but mobile sources (i.e., users); our target becomes estimating the sound sources rather than nodes. We will use the following abbreviations in this paper: DSM refers to baseline \cite{gburrek2021geometry}, iDSM refers to infrastructure-based acoustic localization \cite{gburrek2021geometry}, and aSLAM for acoustic-SLAM \cite{evers2018acoustic}.

As for the conversation extraction, we consider LookOncetoHear \cite{veluri2024look}, target conversation extraction \cite{chen2024target}, and CoS \cite{jenrungrot2020cone} as our baselines. We also replace the relay audio with time-dependent speaker embedding as a baseline.
For LookOncetoHear \cite{veluri2024look}, we assume the enrollment happened in every sentence and takes effect in the next sentence. For the second baseline, we consider self-voice to be the condition. As for the CoS \cite{jenrungrot2020cone}, we assume the interferences span randomly in the nearby space while the target speaker is always in front of the user.
We will use the following abbreviations in this paper: LOTH stands for LookOnceToHear \cite{veluri2024look}, TC refers to \cite{chen2024target}, CoS refers to \cite{jenrungrot2020cone}

\begin{figure}
    \centering
    \begin{subfigure}{0.48\linewidth}
        \centering
\includegraphics[width=0.75\linewidth]{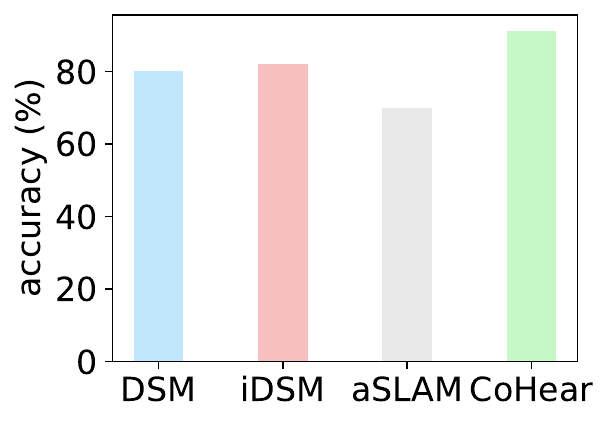}
        \caption{Network topology accuracy.}
        \label{fig:topology accuracy}
    \end{subfigure}%
    \begin{subfigure}{0.48\linewidth}
        \centering
        \includegraphics[width=0.75\linewidth]{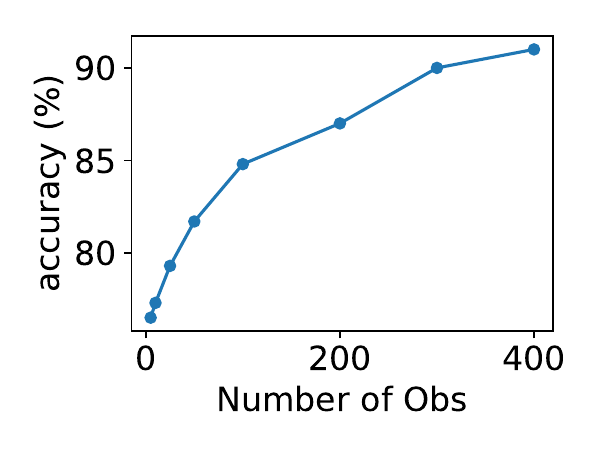}
        \caption{Network topology accuracy with increasing observations.}
        \label{fig:topology accuracy_obs}
    \end{subfigure}%
     \vspace{-1em}
  \caption{Conversation setup performance.}
   \vspace{-1em}
\label{fig:topology}
\end{figure}

\begin{figure}
    \centering
    \begin{subfigure}{0.48\linewidth}
        \centering
        \includegraphics[width=0.75\linewidth]{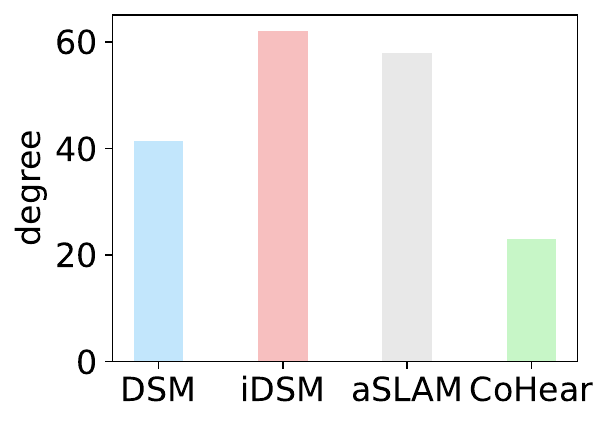}
        \caption{Orientation errors.}
        \label{fig:calibration_output_ori}
    \end{subfigure}%
    \begin{subfigure}{0.48\linewidth}
        \centering
        \includegraphics[width=0.75\linewidth]{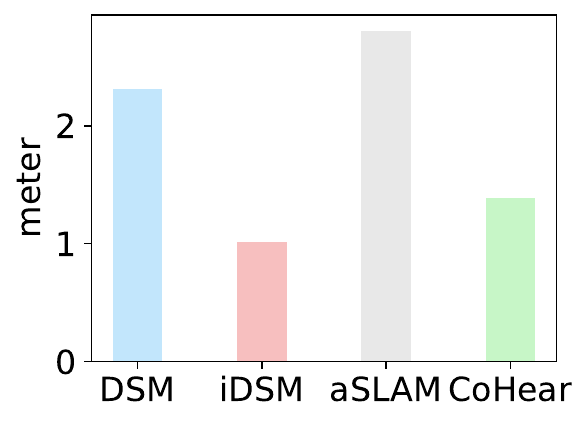}
        \caption{Position errors.}
        \label{fig:calibration_output_pos}
    \end{subfigure}%
      \vspace{-1em}
  \caption{Performance on geometric calibration.}
\label{fig:calibration_output}
\vspace{-1em}
\end{figure}

\subsection{System Performance}

\paragraph{Conversation detection}
We compare \systemname{} with three previously introduced baselines in terms of connection setup, as shown in Fig. \ref{fig:topology accuracy}. Overall, \systemname{} outperforms all the baselines by up to 20\%, achieving over 90\% accuracy in detecting connections. This demonstrates its effectiveness in successfully establishing the network.
In addition, we present the accuracy across different observation counts in Fig. \ref{fig:topology accuracy_obs}. Here, we can see that the accuracy starts at 76\% with just five observations and eventually increases to over 90\% as more observations are included.

\paragraph{Geometric calibration}
We compare \systemname{} with three baselines we introduced before in Fig. \ref{fig:calibration_output}. Overall, \systemname{} outperforms all the baselines except for position errors compared to iDSM. However, iDSM is an infrastructure-based solution with much higher deployment overhead. In summary, \systemname{} reduce the orientation errors up to 43\% and up to 28\% for position errors.

\begin{figure}
    \centering
    \begin{subfigure}{0.48\linewidth}
        \centering
        \includegraphics[width=0.75\linewidth]{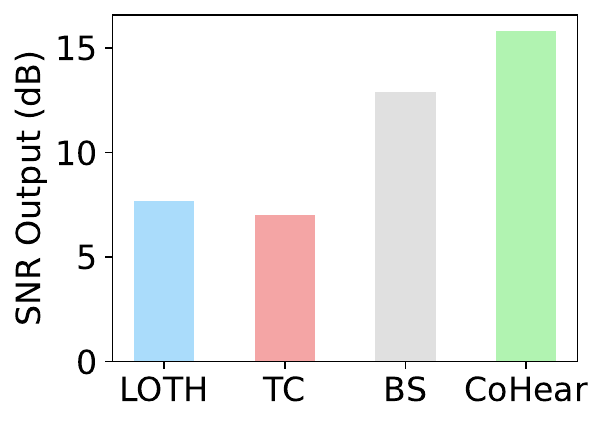}
        \caption{Compared to baselines.}
        \label{fig:compare_speech}
    \end{subfigure}%
    \begin{subfigure}{0.48\linewidth}
        \centering
        \includegraphics[width=0.75\linewidth]{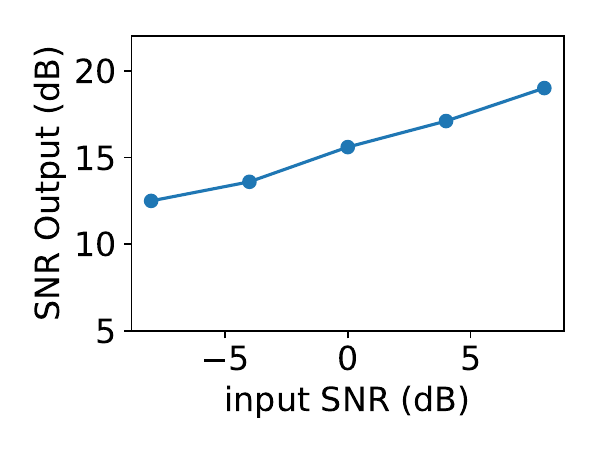}
        \caption{Performance with different input SNR.}
        \label{fig:input_snr}
    \end{subfigure}%
     \vspace{-1em}
    \caption{Performance on conversation extraction, }
    \label{fig:compare}
        \vspace{-1em}
\end{figure}

\paragraph{Conversation extraction}
We compare \systemname{} against the three baselines illustrated before, as shown in Fig. \ref{fig:compare}. We observe that \systemname{} outperforms them with large margin, especially for LOTH (8dB improvement) and TC (8.8dB improvement). As for the CoS, \systemname{} obtains better performance without knowing the location of the speaker, enabling more flexible applications.
The performance of the speech filter also correlates with the input SNR. Apparently, the higher the SNR of the input, the higher the SNR of the output. In the previous evaluation, we already used the difference between input and output to avoid bias. However, it is still interesting to observe the performance under different levels of input SNR in Fig. \ref{fig:input_snr}.

\begin{figure}
    \centering
    \begin{subfigure}{0.48\linewidth}
        \centering
        \includegraphics[width=1\linewidth]{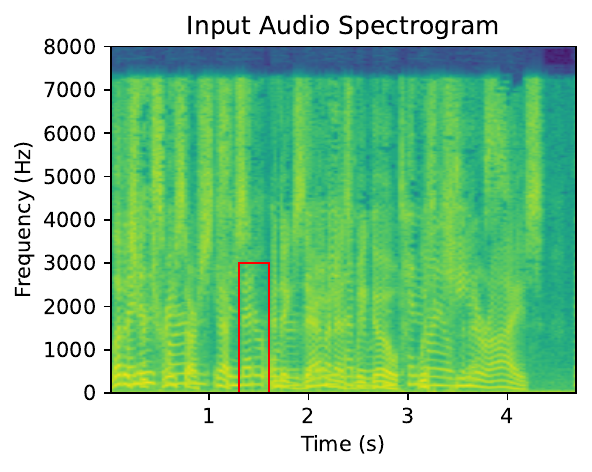}
        \caption{Spectrogram of the input noisy speech.}
        \vspace{-1em}
        \label{fig:visual_input}
    \end{subfigure}%
    \begin{subfigure}{0.48\linewidth}
        \centering
        \includegraphics[width=1\linewidth]{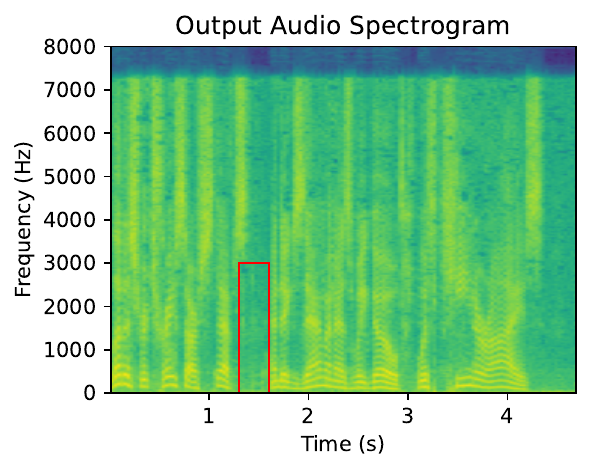}
        \caption{Spectrogram of the output of \systemname{}.}
        \vspace{-1em}
        \label{fig:visual_output}
    \end{subfigure}%
     
    \caption{Spectrograms of speech before and after the processing, the red box highlights the interference which is desired to be removed. The speech content is ``Let us retrace our steps...''.}
    \label{fig:visual}
    \vspace{-1em}
\end{figure}

\paragraph{Visual analysis}
In addition to the results we presented using conventional metrics, we also show the input and output of \systemname{} in Fig. \ref{fig:visual}. In the input spectrogram depicted in Fig. \ref{fig:visual_input}, we notice significant overlap of speech sounds, which occur at nearly the same frequency band. However, with \systemname{}, the interfering speech is effectively removed, as illustrated in Fig. \ref{fig:visual_output}. To enhance clarity, we have highlighted the noise in both figures with a red box, and we can observe that it disappears in the output shown in the right figure.

\paragraph{Ablation Study}
\begin{table}
    \centering
    \caption{Ablation Study for Geometric Calibration}
     \vspace{-1em}
    \begin{tabular}{|l|c|c|}
        \hline
        \textbf{Ablation Component} & \textbf{Position (m)} & \textbf{Orientation (°)} \\
        \hline
        Distance Only & 3.61 & 86.65 \\
        \hline
        Naive Distance & 2.2 & 41. \\
        \hline
        Weight Optimization & 1.69 & 37 \\
        \hline
        \systemname{} & 1.39 & 23.1 \\
        \hline
    \end{tabular}
    \label{tab:ablation_calibration}
        \vspace{-1em}
\end{table}
We conducted an ablation study on geometric calibration by assessing three key components: 1) direction of arrival, 2) collaborative distance estimation, and 3) weight optimization, as shown in Table \ref{tab:ablation_calibration}. Our findings indicate that all three components significantly contribute to the final results. When we used distance estimation as the only observation for calibration, we observed more than a fourfold increase in orientation error. This suggests that relying solely on distance for observations is inadequate for effective calibration.
It is important to note that calibration cannot be achieved with direction of arrival alone, which is why we did not include it in our analysis. Additionally, when we removed collaborative distance estimation, we saw the orientation errors double and position errors increase by 0.8. Lastly, eliminating weight optimization also impacted our results negatively, leading to further degradation in performance.

\subsection{User Study}
\begin{figure}
    \centering
    \begin{minipage}{0.6\linewidth}
        \centering
        \captionof{table}{User study – listening test.}
         \vspace{-1em}
        \begin{tabular}{|c|c|c|}
            \hline
            \textbf{Session} &  \textbf{Score (online)}  & \textbf{Score (offline)}  \\
            \hline
            MOS of Baseline & 2.996 & 3.1 \\
            \hline
            MOS of \systemname{} & 3.9 & 3.7\\
            \hline
            Preferred compression level & 4 & N/A \\
            \hline
            Preferred latency level & 4 & N/A \\
            \hline
        \end{tabular}
        \label{tab:evaluation_ratings1}
    \end{minipage}%
    \hfill
    \begin{minipage}{0.35\linewidth}
        \centering
        \captionof{table}{User study - usability.}
         \vspace{-1em}
        \begin{tabular}{|c|c|}
            \hline
            \textbf{Session} & \textbf{Score} \\
            \hline
            Cannot hear others clearly & 4 \\
            \hline
            Need to look to hear clearly & 4.1  \\
            \hline
            Willing to use earphones & 4.3 \\
            \hline
        \end{tabular}
        \label{tab:evaluation_ratings2}
    \end{minipage}
    \vspace{-1em}
\end{figure}

We conduct a user study evaluation through an online form. We first recruit 20 volunteers to give a subjective evaluation of the enhanced performance of our system online. All of them are college students and did not overlap in the previous data collection. Also, they have no knowledge of our system design before the test. We inform them of the idea of \systemname{} and the aim of the study. 
We follow the ITU P.835 test procedure \cite{gunawan2006subjective} to perform the study. Specifically, the participants are asked to listen to the output of \systemname{} and baseline (without processing) and rate them in random order. The rating is a 5-point scale; the higher the points, the better, and we take the average of them as the mean opinion of score (MOS). The results are shown in Tab. \ref{tab:evaluation_ratings1} and Tab. \ref{tab:evaluation_ratings2}

\paragraph{Session 1: comparison with baseline}
\new{
The first session of the user study evaluates the improvement of \systemname{} against the baseline.
First, each participant is randomly assigned 20 samples for rating (10 by the baseline \cite{veluri2024look} and 10 by \systemname{}).
As shown in Tab. \ref{tab:evaluation_ratings1}, our method has a score more than 1 point higher than the baseline, which means that the participants favor our result.
Moreover, we also conducted a small-scale online study with five participants, where they were instructed to wear our prototype (with ANC headphones) and have a face-to-face talk with others. We also ask another volunteer to act as a competing speaker.
The results from this live study were consistent with the online listening test, which provides strong validation that the perceived audio quality of our enhancement algorithm translates effectively to a real-time usage scenario.
}

\paragraph{Session 2: compression and latency}
The second session of the user study evaluates the performance of \systemname{} under different compression levels (bandwidths). Different from session one, each participant is assigned multiple audio files in sequence, from low compression rate to high compression rate. They are instructed to identify when they feel the quality is already good for usage. We have five levels of compression (8, 16, 32, 64, 128) in the session, so that the result still lies between 1 and 5. Note that the higher the result is, the better \systemname{} performs under limited bandwidth. Similarly, we also ask the participants about their feelings about different levels of latency. Each participant is assigned multiple audio files in sequence, with five different levels of latency (10ms, 20ms, 50ms, 100ms, 200ms). They are instructed to identify when they feel the quality is already good for usage. Note that the higher the result is, \systemname{} is more robust to network latency. As shown in Tab. \ref{tab:evaluation_ratings1}, the majority of participants feel good with level four of compression and latency, which means the performance is preserved under diverse conditions.

\paragraph{Session 3: usability}
\new{In the last session, we asked the same participants to evaluate whether \systemname{}'s usability in the real world, as shown in Tab. Since we don't change the form factor of headphones but build our system on existing headphones, we skip the question of the wearing experience.}
\begin{itemize}
    \item Do you feel that you can not hear others clearly in a noisy environment?
    \item Do you agree that looking at the people you are talking to can let you hear them clearly?
    \item Are you willing to wear earphones when they can improve the conversation clarity?
\end{itemize}
The rating is also a 5-point scale, with five being the most positive and one being the most negative. The average rating for the three questions is 4, 4.1, and 4.3, respectively, which shows that they agree on the usability of \systemname{} in the future.

\subsection{Impact Factors}
In the following analysis, we will examine the key factors of \systemname{} to evaluate the system comprehensively. As a collaborative system, \systemname{} encompasses factors that impact local data processing as well as those that affect transmitted data.
To provide clarity, the first three subsections will focus on local parameters, which will be assessed through the effectiveness of geometric calibration. The final subsection will then explore factors related to conversation extraction.

\begin{figure}
    \centering
    \begin{subfigure}{0.48\linewidth}
        \centering
        \includegraphics[width=0.75\linewidth]{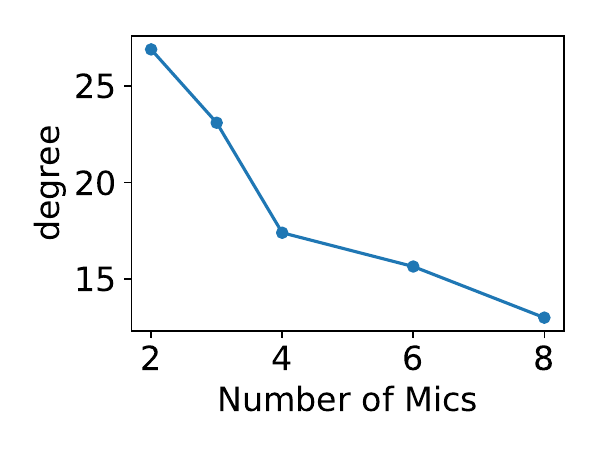}
        \vspace{-1em}
        \caption{Orientation errors.}
        \label{fig:ori_mics}
    \end{subfigure}%
    \begin{subfigure}{0.48\linewidth}
        \centering
        \includegraphics[width=0.75\linewidth]{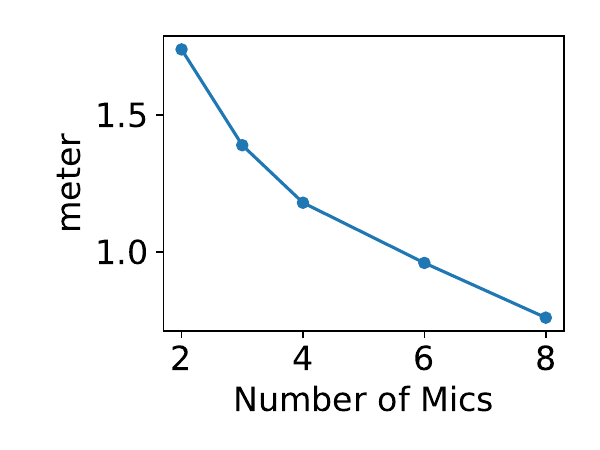}
         \vspace{-1em}
        \caption{Position errors.}
        \label{fig:pos_mics}
    \end{subfigure}%
      \vspace{-1em}
    \caption{Geometric calibration with the number of microphones.}
    \label{fig:micro_mics}
    \vspace{-1em}
\end{figure}

\paragraph{Number of microphones}
The number of microphones has a significant impact on sound localization performance, which also affects the optimization of geometric calibration. To assess this, we conducted tests with different numbers of microphones. As shown in Fig. \ref{fig:micro_mics}, increasing the number of microphones leads to a decrease in estimation errors. Remarkably, even with only two microphones, the orientation estimation error is kept below 30 degrees.

\begin{figure}
    \centering
    \begin{subfigure}{0.48\linewidth}
        \centering
        \includegraphics[width=0.75\linewidth]{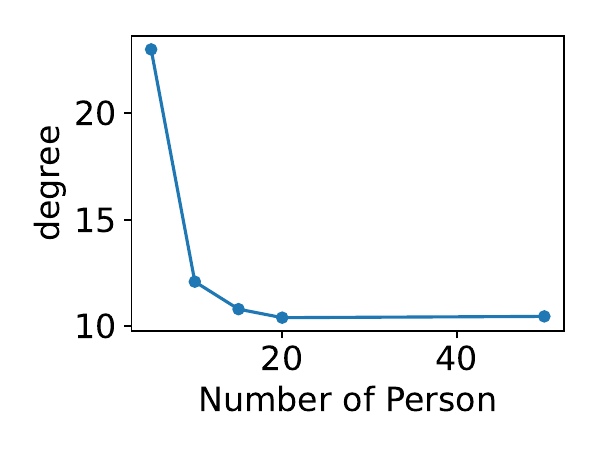}
         \vspace{-1em}
        \caption{Orientation errors.}
        \label{fig:ori_par}
    \end{subfigure}%
    \begin{subfigure}{0.48\linewidth}
        \centering
        \includegraphics[width=0.75\linewidth]{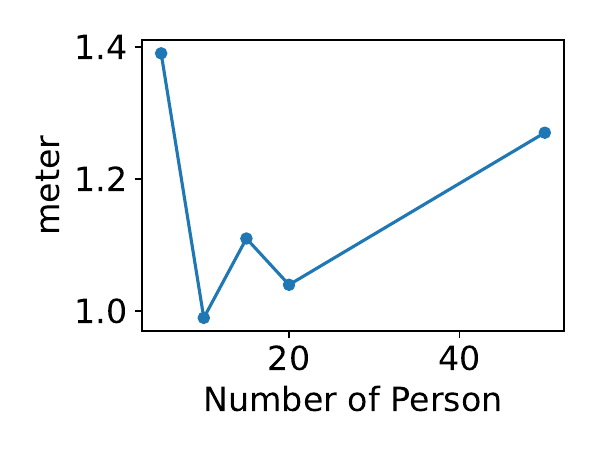}
         \vspace{-1em}
        \caption{Position errors.}
        \label{fig:pos_par}
    \end{subfigure}%
      \vspace{-1em}
    \caption{Geometric calibration with the number of participants.}
    \label{fig:micro_par}
     \vspace{-1em}
\end{figure}

\paragraph{Number of participants}
The number of participants appears to be closely related to the effectiveness of geometric calibration. However, as shown in Fig. \ref{fig:micro_par}, we observe a clear trend indicating that calibration error decreases with the number of participants, especially for the orientation error. Since the orientation accuracy is closely related to conversation accuracy, we conclude that \systemname{} can benefit from more participants.

\begin{figure}
    \centering
    \begin{subfigure}{0.48\linewidth}
        \centering
        \includegraphics[width=0.75\linewidth]{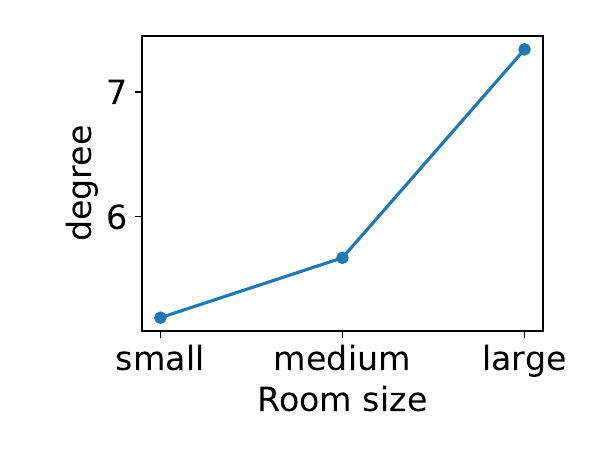}
         \vspace{-1em}
        \caption{Orientation errors.}
        \label{fig:ori_par}
    \end{subfigure}%
    \begin{subfigure}{0.48\linewidth}
        \centering
        \includegraphics[width=0.75\linewidth]{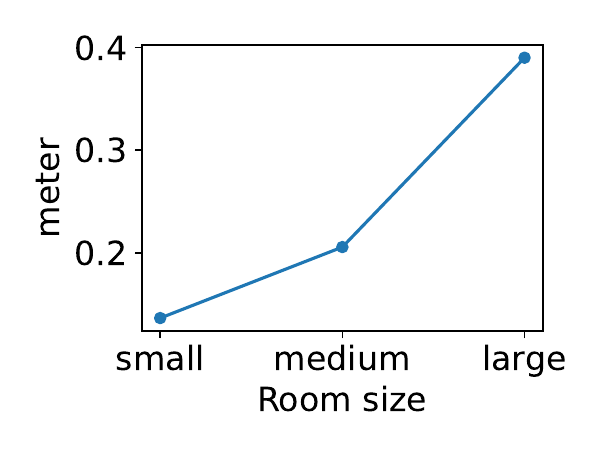}
         \vspace{-1em}
        \caption{Position errors.}
        \label{fig:pos_par}
    \end{subfigure}%
      \vspace{-1em}
    \caption{Geometric calibration with the size of rooms.}
    \label{fig:micro_size}
     \vspace{-1em}
\end{figure}

\paragraph{Size of room}
\new{
The size of the room affects the distance between users (while keeping the number of users constant), which in turn impacts the performance of the geometric calibration. To investigate this, we tested three room sizes: small (5x5 to 10x10 meters), medium (10x10 to 15x15 meters), and large (15x15 to 20x20 meters). As shown in Fig. \ref{fig:micro_size}, the results show a clear trend of increasing error with room size. This matches our expectation, as users positioned closer together lead to better calibration performance.
}

\begin{figure}
    \centering
    \begin{subfigure}{0.48\linewidth}
        \centering
        \includegraphics[width=0.75\linewidth]{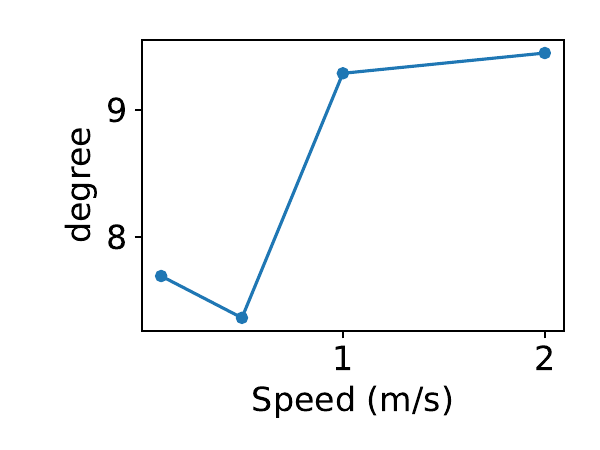}
         \vspace{-1em}
        \caption{Orientation errors.}
        \label{fig:ori_speed}
    \end{subfigure}%
    \begin{subfigure}{0.48\linewidth}
        \centering
        \includegraphics[width=0.75\linewidth]{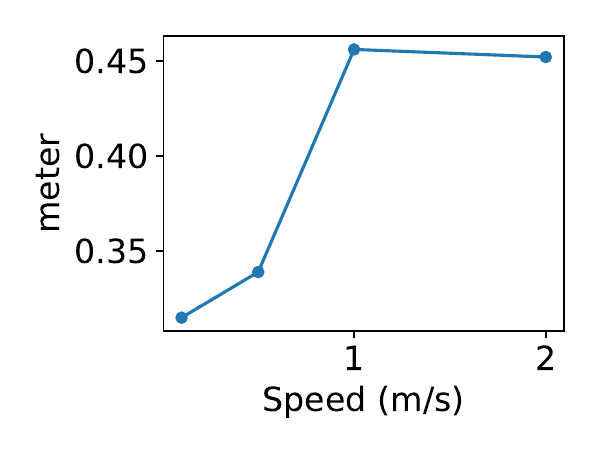}
         \vspace{-1em}
        \caption{Position error.}
        \label{fig:pos_speed}
    \end{subfigure}%
     \vspace{-1em}
    \caption{Geometric calibration with movement speed.}
    \label{fig:micro_speed}
    \vspace{-1em}
\end{figure}

\paragraph{Movement speed}
\new{The movement of participants can affect the geometric calibration by altering the observations. We find that moving speed is a key factor that distinguishes \systemname{} from static WASN, which can be considered as having a speed of 0. In Fig. \ref{fig:micro_speed}, we present the localization errors at various speed levels. Our findings indicate that speed may not have a significant impact on geometric calibration, demonstrating that our system is robust across different speeds.}

\begin{figure}[htbp]
    \centering
    \begin{minipage}{0.45\linewidth}
        \centering
        \includegraphics[width=0.7\linewidth]{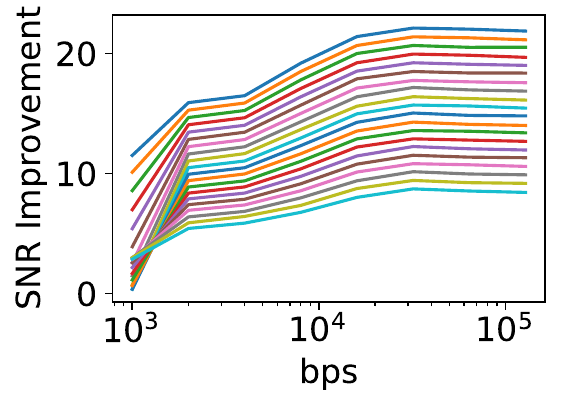}
        \vspace{-1em}
        \caption{Conversation enhancement under different bandwidth, where each line refers to different levels of input SNR.}
        \label{fig:measurement1}
    \end{minipage}
    \hfill
    \begin{minipage}{0.45\linewidth}
        \centering
        \includegraphics[width=0.7\linewidth]{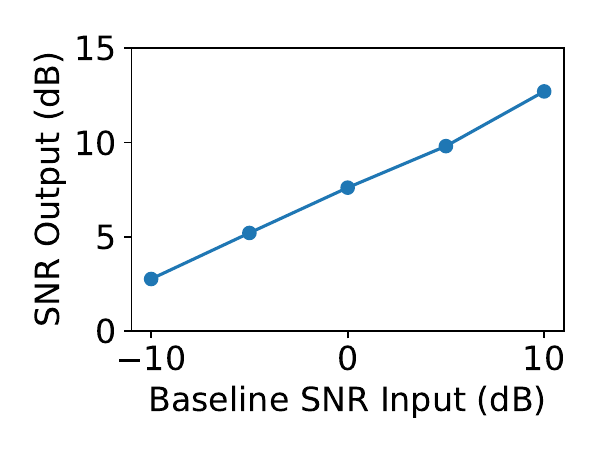}
        \vspace{-1em}
        \caption{Conversation enhancement with time-invariant feature only.}
        \label{fig:baseline_measurement}
    \end{minipage}
    \vspace{-1em}
\end{figure}

\paragraph{Bandwidth}
We evaluate the correlation between performance and compression ratios at different interference levels in Fig. \ref{fig:measurement1}. For each line that refers to one input SNR, we observe that the relationship between bandwidth and performance does not follow a straightforward linear pattern.
To further reduce bandwidth usage, we can rely solely on the speaker embedding when performance is adequate, and we conducted an additional evaluation in Fig. \ref{fig:baseline_measurement}. If the output at 10 dB is considered satisfactory, our observations indicate that enabling the speaker embedding when the input SNR exceeds 5 dB is sufficient, indicating the adaptability of \systemname{} with constrained network.

\begin{figure}
    \centering
    \begin{subfigure}{0.48\linewidth}
        \centering
        \includegraphics[width=0.75\linewidth]{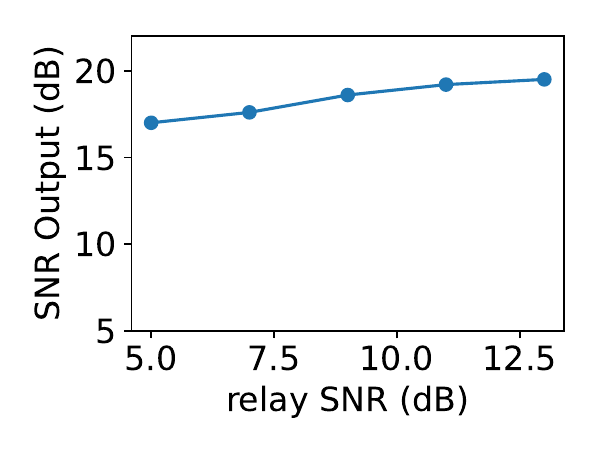}
         \vspace{-1em}
         \caption{}
        \label{fig:relay_snr}
    \end{subfigure}%
    \begin{subfigure}{0.48\linewidth}
        \centering
        \includegraphics[width=0.75\linewidth]{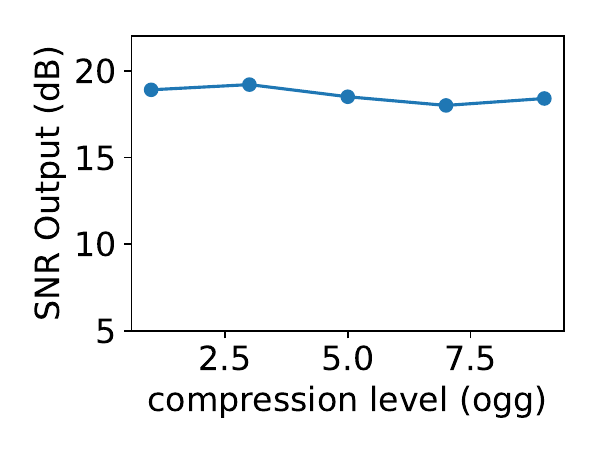}
         \vspace{-1em}
          \caption{}
        \label{fig:relay_compress}
    \end{subfigure}

    \begin{subfigure}{0.48\linewidth}
        \centering
        \includegraphics[width=0.75\linewidth]{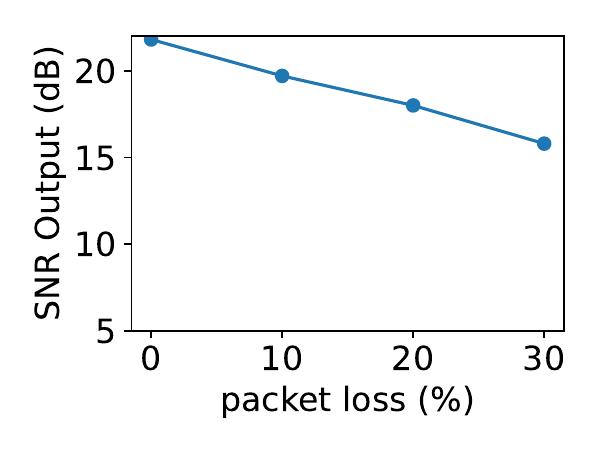}
         \vspace{-1em}
         \caption{}
        \label{fig:relay_packet}
    \end{subfigure}%
    \begin{subfigure}{0.48\linewidth}
        \centering
        \includegraphics[width=0.75\linewidth]{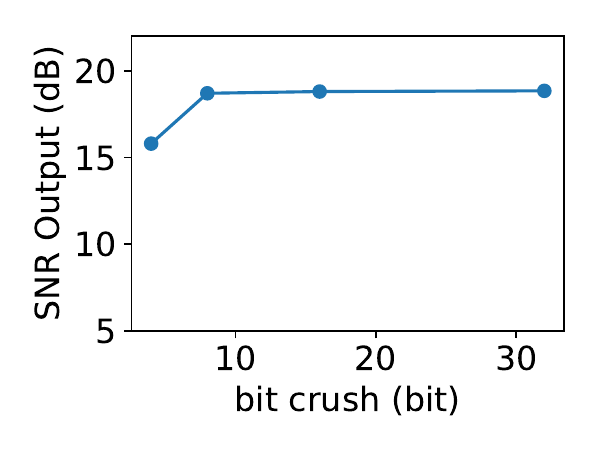}
         \vspace{-1em}
         \caption{}
        \label{fig:relay_bit}
    \end{subfigure}
      \vspace{-1em}
    \caption{Micro-benchmark of conversation extraction under various perturbations on the relay audio.}
    \label{fig:relay_audio_quality}
        \vspace{-1em}
\end{figure}

\paragraph{Relay audio quality}
In the design of \systemname{}, we utilize relay audio as a condition instead of speaker embedding. This choice emphasizes that the quality of the relay audio is critical for the system's performance, which we assess across four aspects, as illustrated in Fig. \ref{fig:relay_audio_quality}.
As shown in Fig. \ref{fig:relay_snr}, there is a positive correlation between the SNR of the relay audio and overall performance. This finding indicates that \systemname{} relies on effective self-voice isolation algorithms, such as VibVoice or ClearBuds \cite{chatterjee2022clearbuds, he2023towards}. Regarding compression quality (Fig. \ref{fig:relay_compress}), we observe that it has only a slight impact on performance, demonstrating that \systemname{} is resilient to compression effects.
The remaining two factors are related to wireless transmission. We note that packet loss negatively affects performance (Fig. \ref{fig:relay_packet}), whereas the bit-crush effect stabilizes at 8 bits (Fig. \ref{fig:relay_bit}). This suggests that we can standardize the transmitted bit width to eight bits without compromising performance.

\begin{figure}
    \centering
    \begin{minipage}{0.48\linewidth}
        \centering
        \includegraphics[width=0.75\linewidth]{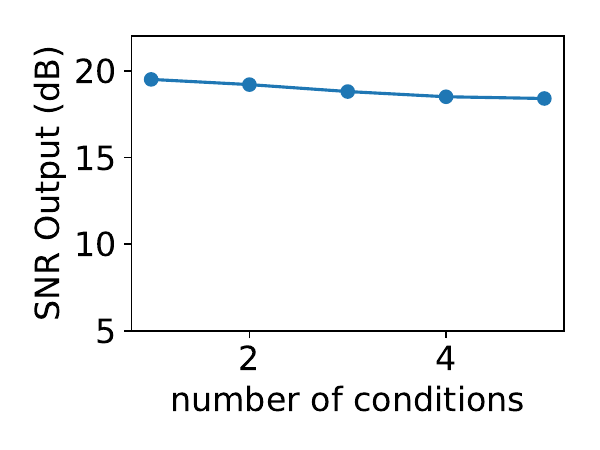}
         \vspace{-1em}
    \caption{Conversation extraction with the number of participants.}
    \label{fig:num_conditions}
    \end{minipage}%
    \hfill
    \begin{minipage}{0.48\linewidth}
        \centering
        \includegraphics[width=0.75\linewidth]{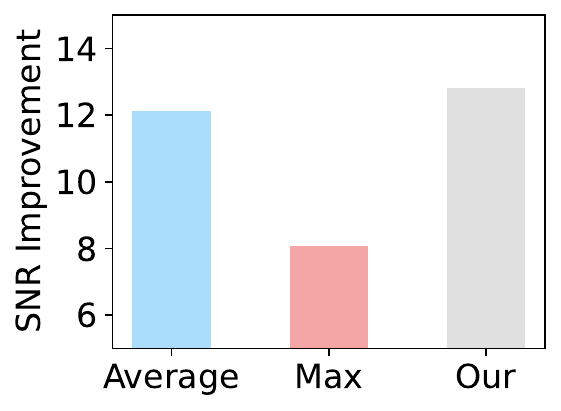}
         \vspace{-1em}
        \caption{Conversation extraction with bandwidth controller.}
        \label{fig:controller}
    \end{minipage}
    \vspace{-1em}
\end{figure}

\paragraph{Number of participants}
Considering that a conversation can involve multiple participants, we also evaluate the performance degradation as the number of speakers increases, as shown in Fig. \ref{fig:num_conditions}. We observe that performance declines slightly with more speakers, which aligns with our intuition.

\subsection{System Overhead}

\begin{figure}
    \centering
    \begin{subfigure}{0.48\linewidth}
        \centering
        \includegraphics[width=0.75\linewidth]{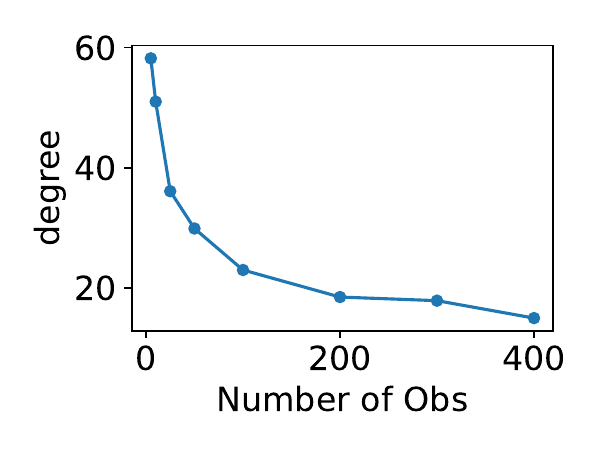}
         \vspace{-1em}
        \caption{Orientation errors.}
        \label{fig:num_obs_orientation}
    \end{subfigure}%
    \begin{subfigure}{0.48\linewidth}
        \centering
        \includegraphics[width=0.75\linewidth]{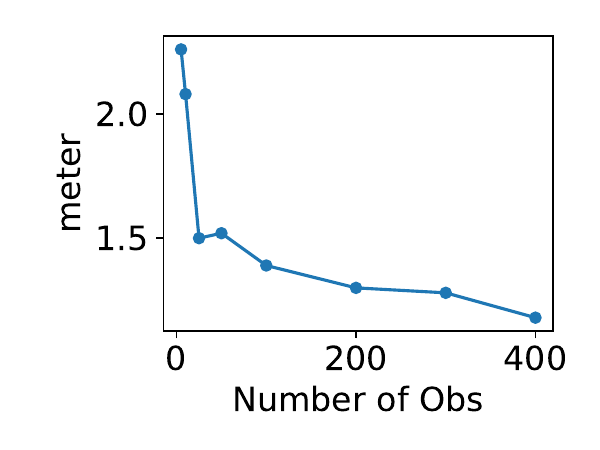}
         \vspace{-1em}
        \caption{Position errors.}
        \label{fig:num_obs_position}
    \end{subfigure}
     \vspace{-1em}
    \caption{Geometric calibration with increasing number of observations.}
    \label{fig:num_obs}
    \vspace{-1em}
\end{figure}

\paragraph{Bandwidth controller}
We assess the performance of our experience-based controller as shown in Fig. \ref{fig:controller}. In this comparison, we evaluate it against two naive baselines: one where each stream receives equal bandwidth, and another where the first stream receives the maximum bandwidth while the second stream receives the remaining bandwidth, so on and so forth. During our experiments, we utilized a random number of nodes, random input SNR, and random available bandwidth. The average SNR is displayed in the figure. Our findings indicate that the proposed controller can achieve an improvement of approximately 1 dB without requiring additional bandwidth.

\paragraph{Cold start time}
\systemname{} depends on a sufficient number of observations or utterances to accurately calibrate the locations of participants. This means that there is a "cold start" period during which the system needs time to achieve optimal performance, as indicated by the conversation accuracy discussed in the previous section.
In Fig. \ref{fig:num_obs}, we present both orientation errors and position errors. Our findings show that \systemname{} requires only a few observations (specifically, around 50) to achieve good results, particularly in orientation estimation.

\paragraph{Runtime latency}
\systemname{} is designed to provide a real-time experience for users, meaning that it is good to have less latency. Since we can process data at the frame level, it is crucial to keep the frame length under 50 ms and ensure that the real-time factor is greater than 1.
We tested the latency on both a smartphone (Pixel 7) and a desktop computer, using either a CPU (i7-11700k) or a GPU (RTX 3090). The results show that the real-time factor (the time to process one frame divided by the duration of the frame) is 2 on the smartphone, 10 on the desktop CPU, and 1250 on the desktop GPU. Compared to baselines such as \cite{veluri2024look}, our system exhibits similar overhead since the majority of the neural network architecture is the same, with only additional computation required for the encoder.

It is also important to note that we do not include the latency of geometric calibration in our real-time inference measurements, as this process can be performed on a server in parallel and takes only 0.2 seconds for 100 observations.

\section{Discussion}
\label{sec:discuss}

\new{
\systemname{} currently uses Wi-Fi for its high throughput and multi-user support, but it is energy-intensive and unsuitable for low-end devices. In contrast, Bluetooth offers low-power, real-time communication but has poor multi-user scalability. Although Bluetooth mesh enables multi-user connections, it provides insufficient bandwidth.
To implement \systemname{} on a portable device without Wi-Fi, we propose two approaches: 1) For future work, developing a custom multi-party audio network over Bluetooth could be explored, as the total theoretical bandwidth is sufficient. 2) Using existing BLE Mesh (which has low bandwidth and high latency), \systemname{} can still function by frequently updating only the speaker embedding, which requires very low data rates.
On the other hand, \systemname{} uses simulated data due to the prohibitive cost and complexity of large-scale real-world experiments with motion capture. Towards an affordable, large-scale data collection, Ultra-Wideband (UWB) presents a conventional, accurate, and budget-friendly alternative worth exploring. Although UWB lacks rich information about the body movement, we are only interested in the position and orientation of users from the context of \systemname{}, which is accessible with UWB.

\paragraph{Multi-modal fusion}
Currently, our geometric calibration only uses IMU data for human tracking in a post-processing manner: we utilize the tracking results from IMU rather than processing its raw data directly. A more advanced approach would be to integrate IMU data into the sound source localization process itself, enabling multi-modal fusion for sound source localization. This integration could compensate for motion-induced errors and potentially extend the system's localization capabilities.

\paragraph{Hardware}
A limitation of the current \systemname{} implementation is its reliance on smartphone computational resources, which introduces undesirable latency and compatibility constraints. A promising future direction is to miniaturize the entire pipeline—including conversation construction and speech enhancement—for direct deployment on a headphone, with the help of the speech AI accelerator \cite{itani2025wireless}. Otherwise, smart glasses represent another viable platform, offering comparatively greater computational power and integrated peripherals like cameras and screens. 

\paragraph{Remote conversation}
While \systemname{} is designed for in-person conversations, its core principles are equally applicable to virtual meetings. The system relies on three key features: head direction, location, and voice. In a remote setting, we can also capture the attention of the participants as well as their voices via webcam and microphone. Although physical location is not directly available, it can be inferred through the participant's position in a virtual space. Consequently, all functionalities of \systemname{} can be effectively replicated in a remote environment, enabling wider applications.
}

\section{Conclusion}
We introduced \systemname{}, a system that enhances voice clarity in noisy environments through multi-device collaboration. By combining advanced acoustic processing, geometric calibration, and conversation enhancement, \systemname{} effectively isolates target conversations in crowded settings.
Our implementation on commercial headphones shows substantial performance improvements over existing solutions. This framework addresses conversation tracking and audio isolation challenges while ensuring scalability and flexibility.
Overall, \systemname{} advances assistive auditory technologies, paving the way for improved communication in complex acoustic environments.

\newpage

\bibliographystyle{ACM-Reference-Format}
\bibliography{sample-base}

\newpage
\appendix

\end{document}